\documentclass[prl,aps,twocolumn,superscriptaddress,a4paper,sort&compress,balancelastpage]{revtex4-2}

\usepackage{amsmath,amssymb,calrsfs}
\usepackage{graphicx}% Include figure files
\usepackage[colorlinks=true,citecolor=blue,urlcolor=blue,linkcolor = blue]{hyperref}  %  For hyperlinks
\usepackage{dcolumn}% needed for some tables
\usepackage{bm}% for math
\usepackage{verbatim}% for math
\usepackage{units}
\usepackage{upgreek}
\usepackage{multirow}

\usepackage[T1]{fontenc}
\usepackage[utf8]{inputenc}

\usepackage[dvipsnames]{xcolor}

\usepackage[normalem]{ulem}

\usepackage{float}

\newcommand{\Sec}[1]{{\textit{#1.---}}}

\newcommand{\TITLE}{Observation of Relaxation Stages in a Nonequilibrium Closed Quantum System: Decaying Turbulence in a Trapped Superfluid}

\begin{document}

\title{\TITLE }

\author{M. A. Moreno-Armijos}
    \email[Corresponding author: ]{michelle.moreno@ifsc.usp.br}
	\affiliation{Instituto de F\'isica de S\~ao Carlos, Universidade de S\~ao Paulo, CP 369, 13560-970 S\~ao Carlos, Brazil}

\author{A. R. Fritsch}
	\affiliation{Instituto de  F\'isica de S\~ao Carlos, Universidade de S\~ao Paulo, CP 369, 13560-970 S\~ao Carlos, Brazil}

\author{A. D. Garc\'{i}a-Orozco}
	\affiliation{Instituto de F\'isica de S\~ao Carlos, Universidade de S\~ao Paulo, CP 369, 13560-970 S\~ao Carlos, Brazil}

\author{S. Sab}
    \affiliation{Instituto de F\'isica de S\~ao Carlos, Universidade de S\~ao Paulo, CP 369, 13560-970 S\~ao Carlos, Brazil}

\author{G. Telles}
	\affiliation{Instituto de F\'isica de S\~ao Carlos, Universidade de S\~ao Paulo, CP 369, 13560-970 S\~ao Carlos, Brazil}

\author{Y. Zhu}
\affiliation{Université Côte d'Azur, CNRS, Institut de Physique de Nice (INPHYNI), 17 rue Julien Lauprêtre 06200 Nice, France}

\author{L. Madeira}
	\affiliation{Instituto de F\'isica de S\~ao Carlos, Universidade de S\~ao Paulo, CP 369, 13560-970 S\~ao Carlos, Brazil}

\author{S. Nazarenko}
\affiliation{Université Côte d'Azur, CNRS, Institut de Physique de Nice (INPHYNI), 17 rue Julien Lauprêtre 06200 Nice, France}

\author{V. I. Yukalov }
	\affiliation{Instituto de F\'isica de S\~ao Carlos, Universidade de S\~ao Paulo, CP 369, 13560-970 S\~ao Carlos, Brazil}
        \affiliation{Bogoliubov Laboratory of Theoretical Physics, Joint Institute for Nuclear Research, Dubna 141980, Russia}

\author{V. S. Bagnato}
	\affiliation{Instituto de F\'isica de S\~ao Carlos, Universidade de S\~ao Paulo, CP 369, 13560-970 S\~ao Carlos, Brazil}
	\affiliation{Department of Biomedical Engineering, Texas A\&M University, College Station, Texas 77843, USA}
        \affiliation{Department of Physics \& Astronomy, Texas A\&M University, College Station, Texas 77843, USA}
	
\date{\today}

\begin{abstract}
The dynamics of nonequilibrium closed quantum systems and their route to thermalization are of fundamental interest to several fields, from cosmology to particle physics. However, a comprehensive description of nonequilibrium phenomena still presents a significant challenge. In this work, we report the observation of distinct stages during the relaxation of the decaying turbulence in trapped Bose-Einstein condensates. Our findings show a direct particle cascade from low to high momenta, a consequence of the energy injection in the system, exhibiting a characteristic universal scaling. This stage is followed by an inverse particle cascade responsible for repopulating the previously depleted condensate. Both cascades can be explained through self-similar solutions provided by wave turbulence theory. These findings provide important insights into the relaxation stages of out-of-equilibrium quantum many-body systems.
\end{abstract}

\maketitle

\Sec{Introduction} Despite the significant progress in our understanding of closed quantum many-body systems, several important challenges persist. One such challenge, present in all areas of physics, concerns thermalization~\cite{Eisert2015,Mori2018}, notably the dynamics of far-from-equilibrium quantum systems and their route to equilibration~\cite{Berges2004,Polkovnikov2011,yukalov2011equilibration,Langen2016}. Experiments with ultracold atoms are well suited to produce and investigate nonequilibrium states due to the high degree of control available,
the flexibility of the excitation protocols, and isolation from the environment~\cite{Rigol2008,trotzky2012probing,gring2012relaxation}, contributing to the rapid advancements on the topic~\cite{Langen2015}. A paradigmatic example of far-from-equilibrium phenomena is quantum turbulence~\cite{Seman2011,Madeira2020}, which has been produced in trapped cold atom systems~\cite{PhysRevLett.103.045301} and displays a characteristic particle cascade~\cite{Thompson_2014,navon2016emergence}. The study of the decaying turbulence regime, which happens once external forces have ceased, has provided important insights into the relaxation dynamics of quantum fluids~\cite{Nowak2012,kwon2014relaxation,Nowak2014,Reeves2022}.

The search for universal features in the dynamics of far-from-equilibrium quantum systems is motivated not only by the possibility of describing complex phenomena in terms of just a few parameters but also by identifying common properties shared by several systems, regardless of the underlying physical mechanisms. The so-called dynamical scaling is one of the most important sought-after properties of dynamic phenomena since it provides a comprehensive framework for understanding the evolution of systems exhibiting self-similarity over time.

Family and Viczek were pioneers in writing an explicit scaling relation with universal exponents~\cite{Famili_1,Famili_2}. The theory of nonthermal fixed points (NTFPs)~\cite{Orioli2015,Schmied2019,Mikheev2023}, conceived in analogy with renormalization group theory in equilibrium critical phenomena, also predicts universal dynamical scaling when a system in a far-from-equilibrium state is in the vicinity of an NTFP. Additionally, the first kind of self-similar solution in the weak wave turbulence theory (WTT)~\cite{nazarenko2011wave} describes how the energy cascades in turbulent systems in a self-similar manner, allowing for the prediction of spectral distributions~\cite{Zhu2023,Zhu2023PRE}. The theories above were developed to explain distinct physical systems with diverse microscopic mechanisms governing their time evolution. However, in all of them, the universal dynamical scaling of the relevant quantity $f(x,t)$ can be cast into the form  $f(x,t) = t^a F(t^b x)$, where $F$ is the universal function and $a$ and $b$ are the universal exponents.

\begin{figure*}[t]
    \centering
    \includegraphics[scale = 1]{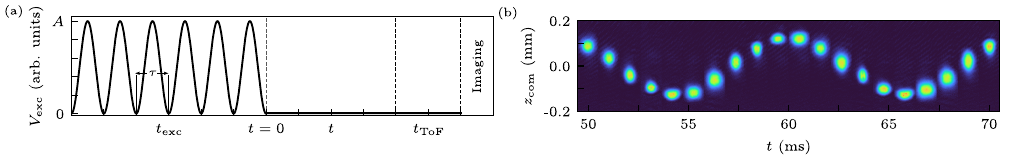}
    \caption{(a) Representation of the experimental protocol. After producing a BEC, an external time-dependent oscillatory potential with amplitude $A$ and period $\tau$ is applied during excitation time $t_{\mathrm{exc}}$. At $t=0$, the external potential is turned off, and the cloud evolves in the trap during a hold time $t$. Then, the cloud is released, and an absorption imaging is taken after 30 ms of time of flight $t_{\mathrm{TOF}}$. (b) Absorption images for an excitation amplitude of $A = 2.55\mu_0$ showing the center-of-mass position $z_{\rm com}$ along the $z$ axis in the time interval $50<t<70$ ms, evidencing the dipolar and quadrupolar modes.}
    \label{fig:excitation_dipolar}
\end{figure*}

In this work, we observe distinct stages in the relaxation dynamics of a harmonically trapped three-dimensional Bose-Einstein condensate (BEC) driven to a turbulent state. We begin with the condensate in equilibrium and inject energy into the system through a controllable excitation protocol. After we cease the external driving, we observe universal dynamical scaling in the time evolution of the momentum distribution $n(k,t)$, which takes the form
\begin{equation}
    \label{eq:scaling_momentum_dist}
    n(k,t) = (t/t_{\mathrm{ref}})^{\alpha} n\left[ (t/t_{\mathrm{ref}})^{\beta} k,t_{\rm ref} \right].
\end{equation}
Here, $\alpha$ and $\beta$ are the universal exponents, $t$ is the elapsed time since the relaxation started, and $t_{\mathrm{ref}}$ is an arbitrary reference time within the temporal window when scaling is observed. This universal dynamics is associated with a direct energy cascade from the low- to the high-momentum region, which depletes the condensate~\cite{PhysRevA2022}. By controlling the amount of energy given to the system, we observe that after the direct cascade, there is the appearance of an inverse cascade responsible for repopulating the condensate, which follows the dynamical scaling
\begin{equation}
\label{eq:ss2nd}
n(k,t) = \tau^{\lambda} n(\tau^{\mu} k,t_{\rm ref}), \, \tau=(t_{b}-t)/(t_b-t_{\rm ref}),
\end{equation}
where the condensate population diverges at $t_b$, and the universal exponents are $\mu$ and $\lambda<0$. Observing the depletion and repopulation of the condensate as stages of the relaxation dynamics of the decaying turbulence regime in trapped BECs is an important step toward understanding the thermalization processes in these far-from-equilibrium systems.

\Sec{Experimental method}
We begin by producing a $^{87}\mathrm{Rb}$ BEC in the ground state $F = 1$, $m_F = -1$ of the hyperfine structure, in a magnetic trap characterized by radial and axial frequencies $\omega_r = 2\pi \times 93.0(5)$ Hz and $\omega_x = 2\pi \times 13.5(3)$ Hz, respectively. After completing radio-frequency evaporation, we produce a cigar-shaped BEC with $ N \approx 2 \times 10^5$ atoms, a condensed fraction above $80 \%$, and temperature $T\approx 50$ nK. After we produce the BEC in equilibrium, we begin the excitation protocol as represented in Fig.~\ref{fig:excitation_dipolar}(a). For a time $t_{\mathrm{exc}}$, we apply a time-oscillating magnetic field gradient generated by a pair of coils in an anti-Helmholtz configuration whose symmetry axis is misaligned with respect to the trap center. The applied potential with excitation amplitude $A$, which we present in units of the chemical potential $\mu_0$ of the initial equilibrium BEC, creates deformations and rotations in the trapping potential, exciting the dipolar and quadrupolar modes of the BEC, as shown in Fig.~\ref{fig:excitation_dipolar}(b). After the excitation, we switch off the excitation potential and let the cloud evolve in the trap for a time $t$ that lasts up to 500 ms. We then turn off the trap and let the cloud expand during a time of flight $t_{\mathrm{TOF}} = 30$ ms before taking an absorption image along the $x$ axis that we use to obtain the angular averaged momentum distribution $n(k,t)$~\cite{SupMat}.
The two-dimensional momentum distributions are normalized such that $2\pi \int dk \ k\ n(k,t)=1$, and the total number of atoms is conserved throughout the dynamics~\cite{SupMat}.

\begin{figure*}[ht] 
\centering
\includegraphics[scale=1]{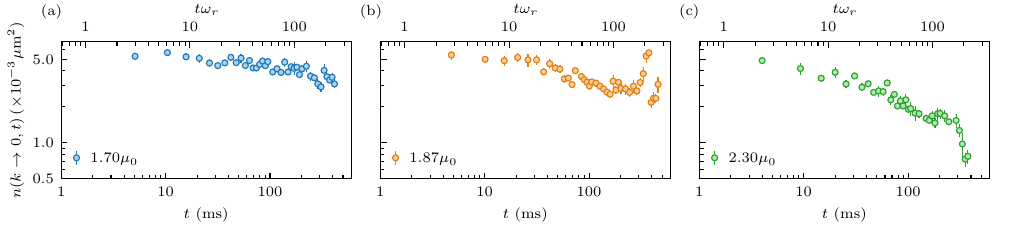}
\caption{Time evolution of $n(k\rightarrow 0,t)$ for three different excitation amplitudes. (a) For $A = 1.70 \mu_0$, the system exhibits a slow decrease in $n(k \rightarrow 0, t)$, corresponding to a depletion of the low-momentum modes. (b) For $A = 1.87\mu_0$, a time interval with an increase of $n(k\rightarrow 0,t)$ is evident, indicating an inversion in the direction of the particle flux and the repopulation of the condensate. (c) For higher values of the excitation amplitude, for instance, $A = 2.30\mu_0$, the repopulation of the low-momentum modes is significantly suppressed. The error bars represent one standard deviation~\cite{SupMat}.}
\label{fig:nk0_amps}
\end{figure*}

\Sec{Evolution of $n(k\rightarrow 0,t)$ during the relaxation} Since we are interested in observing variations in the condensate population, we focus on the low-momentum limit of the momentum distribution, $n(k \rightarrow 0, t)$. Figure~\ref{fig:nk0_amps} shows the time evolution of $n(k \rightarrow 0, t)$ for three excitation amplitudes that illustrate the effect of the external perturbation. For all excitation amplitudes, the relaxation process starts with a decrease in $n(k \rightarrow 0, t)$, indicating a continuous depletion of the low-momentum modes. 
In the case of $A = 1.70 \mu_0$, Fig.~\ref{fig:nk0_amps}(a),
depletion is the only feature present in $n(k \rightarrow 0, t)$. The excitation amplitude is so low we do not observe characteristics of a turbulent system, such as a power-law behavior in the momentum distribution.  For $A = 1.87 \mu_0$, shown in Fig.~\ref{fig:nk0_amps}(b), an

\onecolumngrid

\newpage

\twocolumngrid

\noindent increase in $n(k \rightarrow 0, t)$ follows the depletion of the condensate. This inversion in the direction of the particle flux is responsible for the repopulation of the condensate. Higher excitation amplitudes deplete the condensate faster due to more energy being injected into the system. This is the case of $A = 2.30\mu_0$, Fig.~\ref{fig:nk0_amps}(c), where the repopulation of the condensate does not occur.

\begin{figure}[htb]
    \centering
    \includegraphics[scale=1]{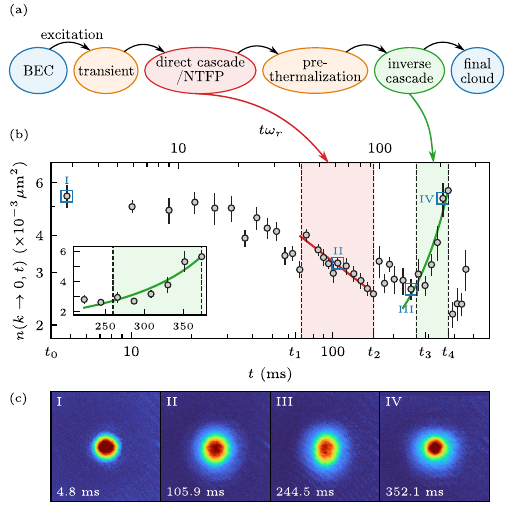}
    \caption{Relaxation stages identified by examining the evolution of $n(k\rightarrow 0,t)$. (a) Diagram of the sequence of relaxation stages. (b) Low-momentum limit of the momentum distribution $n(k \rightarrow 0,t)$ for $A=1.87\mu_0$. The instant $t_0=0$ indicates the end of the excitation and the beginning of relaxation. The first stage, from $t_0 = 0$ to $t_1=70$ ms, shows a slow and gradual decrease, indicating a direct flow of low-momentum particles to higher momenta. In the second stage, from $t_1$ to $t_2=160$ ms, the system is close to an NTFP and exhibits universal scaling in the form $n(k\rightarrow 0,t)\propto t^\alpha$ (solid red line). We observe a transient in the interval from $t_2$ to $t_3=260$ ms, characterized by small variations in $n(k\rightarrow 0,t)$, indicating a possible prethermalization of the system. This is followed by an inverse particle cascade, reaching a peak at $t_4=370$ ms, showing the repopulation of the condensed mode (more detail in the inset). The green curve corresponds to the low-momentum limit of Eq.~(\ref{eq:ss2nd}), $n(k\to 0,t)\propto (t_{b}-t)^{\lambda}$, where we set the exponent to its WTT prediction, $\lambda=-1.46$, and fit $t_b$.
    The error bars represent one standard deviation~\cite{SupMat}.
    (c) Absorption images illustrate the state of the momentum distribution at a specific time for each of the discussed stages.}
    \label{fig:example_nk_0}
\end{figure}

Now, we concentrate on the relaxation stages that occur when the repopulation of the condensate is present. Our analysis will be based on the $A = 1.87 \mu_0$ excitation amplitude [shown in Fig.~\ref{fig:nk0_amps}(b)], which serves as an illustrative example for other amplitudes displaying low-momenta repopulation. Figure~\ref{fig:example_nk_0}(a) shows a diagrammatic representation of the relaxation stages, corresponding to the features of $n(k\rightarrow 0,t)$, Fig.~\ref{fig:example_nk_0}(b). Some illustrative absorption images are presented in Fig.~\ref{fig:example_nk_0}(c).

The observations begin at $t_0 = 0$, which marks the end of the excitation and the start of the relaxation. We initially observe a slow decrease of $n(k\rightarrow 0,t)$, characterizing a direct flux of particles from low to high momentum. In the interval from $t_1=70$ ms to $t_2=160$ ms, the system approaches an NTFP and thus displays the universal scaling of Eq.~(\ref{eq:scaling_momentum_dist}). Since we are considering only the $k\to 0$ component, the time-dependent scaling depends only on one of the universal exponents, $n(k \rightarrow 0, t) \propto t^\alpha$~\cite{madeira2023differential}, from which we obtain $\alpha=-0.52(6)$. If we consider a finite momentum range, instead of just the $k\to 0$ component, we obtain $\alpha = -0.5(1)$ and $\beta = -0.25(7)$~\cite{SupMat}, in accordance with a previous investigation~\cite{PhysRevA2022}. A transient takes place from $t_2$ to $t_3 = 260$ ms after the direct particle cascade occurs and before the inverse cascade is observed. The fact that $n(k \rightarrow 0, t)$ is approximately constant indicates that we could be observing prethermalization in the system, where the populations of the momentum modes are quasistationary~\cite{Berges2004,Langen2016}. This region of small variation is followed by an inverse cascade when $n(k\rightarrow 0,t)$ increases with time as the system repopulates the condensate, reaching a peak at $t_4 = 370$ ms. This stage displays the dynamical scaling of Eq.~(\ref{eq:ss2nd}), i.e., its low-momentum limit $n(k\to 0,t)\propto (t_{b}-t)^{\lambda}$.
This phenomenon is often called the blowup condensation since at $t_b$, the $k\to 0$ component diverges to infinity~\cite{Zhu2023,Zhu2023PRE}. Our data present a similar feature, indicating this might be responsible for the fast repopulation of the condensate. After the condensate is reestablished, we observe a sudden drop in $n(k \rightarrow 0,t)$~\cite{Temperature}.

\begin{figure*}[htb]
    \centering
    \includegraphics[scale=1]{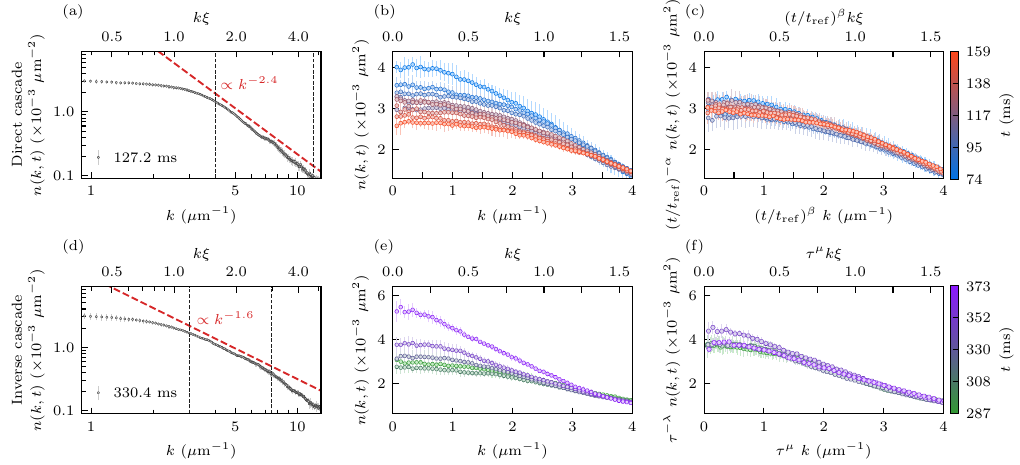}
    \caption{
The appearance of direct (a) and inverse (d) particle cascades during the relaxation for $A = 1.87\mu_0$. The red dashed lines guide the eye regarding the power laws observed in the cascades, their range delimited by the vertical dashed lines. Momentum distributions $n(k,t)$ in the interval where the direct (b) and inverse (e) cascades occur. Two kinds of dynamical scaling are observed during the relaxation, evidenced by the collapse into universal functions. (c) The first is due to a direct particle cascade, which is described by Eq.~(\ref{eq:scaling_momentum_dist}) with the universal exponents $\alpha = -0.5(1)$ and $\beta=-0.25(7)$. (f) The second, associated with an inverse cascade, follows Eq.~(\ref{eq:ss2nd}) with $\lambda=-1.5(5)$ and $\mu=-0.9(3)$. The error bars correspond to one standard deviation~\cite{SupMat}.
}
\label{fig:example_nk_scaling}
\end{figure*}

\Sec{Particle cascades and dynamical scaling} One of the challenges in studying quantum turbulence is that distinct physical mechanisms produce different types of turbulence~\cite{barenghi2023types}. In numerical simulations, access to the amplitude and phase of the wave function allows the identification of the type of turbulence more straightforwardly. Our experiment provides us with the momentum distribution as the main observable; hence, we must infer the underlying physical processes from $n(k,t)$. Our excitation protocol introduces many excitations in the system, including density waves. Therefore, the WTT~\cite{nazarenko2011wave} may be used to describe how waves interact via weakly nonlinear processes. In particular, the WTT~\cite{nazarenko2011wave} provides a wave kinetic equation (WKE) that describes the process of four-wave mixing and predicts self-similar solutions to it.

For the direct energy cascade, the stationary Kolmogorov-Zakharov solution of the WKE yields a $n(k) \propto k^{-2}$ dependence, where we omit the log correction to the marginally divergent integral in the WKE~\cite{Zhu2023,Zhu2023PRE} since it would not be significant in the momentum and excitation amplitude ranges of this investigation. During the direct cascade stage, we observe the momentum distributions close to $\propto k^{-2.4}$, as illustrated in Fig.~\ref{fig:example_nk_scaling}(a). In addition, the so-called self-similar solution of the first kind~\cite{Zhu2023,Zhu2023PRE} describes a direct energy cascade with a dynamical scaling that can be cast into the form of Eq.~(\ref{eq:scaling_momentum_dist}). The prediction for an infinite and homogeneous system is $\alpha=-2/3$ and $\beta=-1/6$~\cite{Zhu2023PRE}. The fact that we have an inhomogeneous finite system may account for the small deviations we observe, $\alpha = -0.5(1)$ and $\beta = -0.25(7)$. The collapse of all momentum distributions in the direct cascade stage, Fig.~\ref{fig:example_nk_scaling}(b), into a universal function for a finite momentum range is depicted in Fig.~\ref{fig:example_nk_scaling}(c). Interestingly, the dynamical scaling provided by the self-similar solution of the first kind is of the same form as the one predicted by the theory of NTFPs~\cite{Orioli2015,Schmied2019,Mikheev2023}, which has been observed in several cold atom experiments~\cite{Nicklas2015observationscaling, prufer2018observation, erne2018universal, eigen2018universal,glidden2021bidirectional,PhysRevA2022,lannig2023observation,galka2022emergence}.

For the inverse particle cascade, WTT predicts a self-similar solution of the second kind, Eq.~(\ref{eq:ss2nd}), with $n(k) \propto k^{-1.52}$ at the larger $k$ side~\cite{Zhu2023,Zhu2023PRE}, and we observe $n(k)\propto k^{-\nu}$ with $\nu=1.6$, as seen in Fig.~\ref{fig:example_nk_scaling}(d). The momentum distributions during the inverse cascade stage, Fig.~\ref{fig:example_nk_scaling}(e), collapse into a universal function, Fig.~\ref{fig:example_nk_scaling}(f), after the dynamical scaling of Eq.~(\ref{eq:ss2nd}) is applied. We obtained $\lambda=-1.5(5)$ and $\mu=-0.9(3)$~\cite{SupMat}, which can be compared with the exponents predicted by WTT, ${\lambda} ={\nu/(2-2\nu)}\approx -1.46$ and $\mu$ ranging from $-1.14$ to $-1.04$~\cite{Zhu2023,Zhu2023PRE}. It is worth noting that WTT was developed considering infinite homogeneous systems, so small deviations between its predicted values and our observations are expected due to our inhomogeneous system with different boundary conditions.

The second kind of self-similar solution predicted by WTT meets the definition of an NTFP; however, its functional form differs from Eq.~(\ref{eq:scaling_momentum_dist}). The fact that this dynamical scaling appears in the dynamics of far-from-equilibrium systems suggests it could be merged into the theory as an NTFP of the second kind, which opens up the possibility of describing a broader class of phenomena using the same framework.

\Sec{Conclusion} In this work, we have observed that the relaxation of a turbulent BEC undergoes various stages en route to thermalization. By tracking the time evolution of the low-momentum limit of the momentum distribution $n(k\rightarrow 0,t)$, we could identify and characterize several of the processes involved in the dynamics. After the end of the external perturbation, the condensate is slowly depleted until it reaches a stage where we observe a direct particle cascade, consistent with the system being close to an NTFP with universal exponents $\alpha = -0.5(1)$ and $\beta = -0.25(7)$. This universal scaling can be explained by the four-wave mixing process in the  WTT~\cite{nazarenko2011wave}. After this stage, we observe a region where $n(k\rightarrow 0,t)$ experiences small variations, indicating a possible prethermalization. Next, an inverse particle cascade takes place in the dynamics of the system, which can be described by another type of self-similar solution to the kinetic equations of the WTT, which is associated with the repopulation of the condensate. For this other type of dynamical scaling, we obtain $\lambda=-1.5(5)$ and $\mu=-0.9(3)$.

Observing the relaxation stages of a turbulent trapped superfluid is an important benchmark for investigating the fundamental process accounting for thermalization in these quantum many-body systems driven far from equilibrium. However, several aspects still need to be investigated, such as the necessary conditions for all stages to occur. Moreover, while our findings show slight differences from the results for infinite homogeneous systems, examining the confining potential and anisotropy effects on the relaxation dynamics warrants further investigation. Remarkably, the WTT provides two types of self-similar solutions that explain the observed direct and inverse cascades using the same fundamental physical mechanism. This indicates that far-from-equilibrium systems may display more than one type of dynamical scaling, even if the microscopical origin of the self-similarity is the same, opening up the possibility of explaining several processes in terms of a single underlying mechanism.

\begin{acknowledgments}
\Sec{Acknowledgments} We thank M.A.~Caracanhas and H.~Perrin for the fruitful discussions. This work was supported by the São Paulo Research Foundation (FAPESP) under Grants No. 2013/07276-1, No. 2014/50857-8, No. 2022/00697-0, and No. 2023/04451-9, by the National Council for Scientific and Technological Development (CNPq) under Grants No. 465360/2014-9 and No. 381381/2023-4, by the EU Marie Skłodowska-Curie RISE project HALT, Grant No. 823937, and ANR VORTEX Grant No. ANR-22-CE30-0011. M.A.M-A. and S.~S. acknowledge the support from Coordenação de Aperfeiçoamento de Pessoal de Nível Superior-Brasil (CAPES) - Finance Codes No. 88887.643259/2021-00 and No. 88887.980579/2024-00. Texas A\&M University (GURI-Governor's University Research Initiative - M230930).
\end{acknowledgments}

% \bibliographystyle{revtex4-1}
%\bibliography{references.bib}

\begin{thebibliography}{39}%
	\makeatletter
	\providecommand \@ifxundefined [1]{%
		\@ifx{#1\undefined}
	}%
	\providecommand \@ifnum [1]{%
		\ifnum #1\expandafter \@firstoftwo
		\else \expandafter \@secondoftwo
		\fi
	}%
	\providecommand \@ifx [1]{%
		\ifx #1\expandafter \@firstoftwo
		\else \expandafter \@secondoftwo
		\fi
	}%
	\providecommand \natexlab [1]{#1}%
	\providecommand \enquote  [1]{``#1''}%
	\providecommand \bibnamefont  [1]{#1}%
	\providecommand \bibfnamefont [1]{#1}%
	\providecommand \citenamefont [1]{#1}%
	\providecommand \href@noop [0]{\@secondoftwo}%
	\providecommand \href [0]{\begingroup \@sanitize@url \@href}%
	\providecommand \@href[1]{\@@startlink{#1}\@@href}%
	\providecommand \@@href[1]{\endgroup#1\@@endlink}%
	\providecommand \@sanitize@url [0]{\catcode `\\12\catcode `\$12\catcode
		`\&12\catcode `\#12\catcode `\^12\catcode `\_12\catcode `\%12\relax}%
	\providecommand \@@startlink[1]{}%
	\providecommand \@@endlink[0]{}%
	\providecommand \url  [0]{\begingroup\@sanitize@url \@url }%
	\providecommand \@url [1]{\endgroup\@href {#1}{\urlprefix }}%
	\providecommand \urlprefix  [0]{URL }%
	\providecommand \Eprint [0]{\href }%
	\providecommand \doibase [0]{https://doi.org/}%
	\providecommand \selectlanguage [0]{\@gobble}%
	\providecommand \bibinfo  [0]{\@secondoftwo}%
	\providecommand \bibfield  [0]{\@secondoftwo}%
	\providecommand \translation [1]{[#1]}%
	\providecommand \BibitemOpen [0]{}%
	\providecommand \bibitemStop [0]{}%
	\providecommand \bibitemNoStop [0]{.\EOS\space}%
	\providecommand \EOS [0]{\spacefactor3000\relax}%
	\providecommand \BibitemShut  [1]{\csname bibitem#1\endcsname}%
	\let\auto@bib@innerbib\@empty
	%</preamble>
	\bibitem [{\citenamefont {Eisert}\ \emph {et~al.}(2015)\citenamefont {Eisert},
		\citenamefont {Friesdorf},\ and\ \citenamefont {Gogolin}}]{Eisert2015}%
	\BibitemOpen
	\bibfield  {author} {\bibinfo {author} {\bibfnamefont {J.}~\bibnamefont
			{Eisert}}, \bibinfo {author} {\bibfnamefont {M.}~\bibnamefont {Friesdorf}},\
		and\ \bibinfo {author} {\bibfnamefont {C.}~\bibnamefont {Gogolin}},\
	}\bibfield  {title} {\bibinfo {title} {Quantum many-body systems out of
			equilibrium},\ }\href {https://doi.org/10.1038/nphys3215} {\bibfield
		{journal} {\bibinfo  {journal} {Nat. Phys.}\ }\textbf {\bibinfo {volume}
			{11}},\ \bibinfo {pages} {124} (\bibinfo {year} {2015})}\BibitemShut
	{NoStop}%
	\bibitem [{\citenamefont {Mori}\ \emph {et~al.}(2018)\citenamefont {Mori},
		\citenamefont {Ikeda}, \citenamefont {Kaminishi},\ and\ \citenamefont
		{Ueda}}]{Mori2018}%
	\BibitemOpen
	\bibfield  {author} {\bibinfo {author} {\bibfnamefont {T.}~\bibnamefont
			{Mori}}, \bibinfo {author} {\bibfnamefont {T.~N.}\ \bibnamefont {Ikeda}},
		\bibinfo {author} {\bibfnamefont {E.}~\bibnamefont {Kaminishi}},\ and\
		\bibinfo {author} {\bibfnamefont {M.}~\bibnamefont {Ueda}},\ }\bibfield
	{title} {\bibinfo {title} {Thermalization and prethermalization in isolated
			quantum systems: a theoretical overview},\ }\href
	{https://doi.org/10.1088/1361-6455/aabcdf} {\bibfield  {journal} {\bibinfo
			{journal} {J. Phys. B}\
		}\textbf {\bibinfo {volume} {51}},\ \bibinfo {pages} {112001} (\bibinfo
		{year} {2018})}\BibitemShut {NoStop}%
	\bibitem [{\citenamefont {Berges}\ \emph {et~al.}(2004)\citenamefont {Berges},
		\citenamefont {Borsanyi},\ and\ \citenamefont {Wetterich}}]{Berges2004}%
	\BibitemOpen
	\bibfield  {author} {\bibinfo {author} {\bibfnamefont {J.}~\bibnamefont
			{Berges}}, \bibinfo {author} {\bibfnamefont {S.}~\bibnamefont {Borsanyi}},\
		and\ \bibinfo {author} {\bibfnamefont {C.}~\bibnamefont {Wetterich}},\
	}\bibfield  {title} {\bibinfo {title} {Prethermalization},\ }\href
	{https://doi.org/10.1103/PhysRevLett.93.142002} {\bibfield  {journal}
		{\bibinfo  {journal} {Phys. Rev. Lett.}\ }\textbf {\bibinfo {volume}
			{93}},\ \bibinfo {pages} {142002} (\bibinfo {year} {2004})}\BibitemShut
	{NoStop}%
	\bibitem [{\citenamefont {Polkovnikov}\ \emph {et~al.}(2011)\citenamefont
		{Polkovnikov}, \citenamefont {Sengupta}, \citenamefont {Silva},\ and\
		\citenamefont {Vengalattore}}]{Polkovnikov2011}%
	\BibitemOpen
	\bibfield  {author} {\bibinfo {author} {\bibfnamefont {A.}~\bibnamefont
			{Polkovnikov}}, \bibinfo {author} {\bibfnamefont {K.}~\bibnamefont
			{Sengupta}}, \bibinfo {author} {\bibfnamefont {A.}~\bibnamefont {Silva}},\
		and\ \bibinfo {author} {\bibfnamefont {M.}~\bibnamefont {Vengalattore}},\
	}\bibfield  {title} {\bibinfo {title} {Colloquium : Nonequilibrium dynamics
			of closed interacting quantum systems},\ }\href
	{https://doi.org/10.1103/RevModPhys.83.863} {\bibfield  {journal} {\bibinfo
			{journal} {Rev. Mod. Phys.}\ }\textbf {\bibinfo {volume} {83}},\
		\bibinfo {pages} {863} (\bibinfo {year} {2011})}\BibitemShut {NoStop}%
	\bibitem [{\citenamefont {Yukalov}(2011)}]{yukalov2011equilibration}%
	\BibitemOpen
	\bibfield  {author} {\bibinfo {author} {\bibfnamefont {V.}~\bibnamefont
			{Yukalov}},\ }\bibfield  {title} {\bibinfo {title} {Equilibration and
			thermalization in finite quantum systems},\ }\href
	{https://doi.org/https://doi.org/10.1002/lapl.201110002} {\bibfield
		{journal} {\bibinfo  {journal} {Laser Phys. Lett.}\ }\textbf {\bibinfo
			{volume} {8}},\ \bibinfo {pages} {485} (\bibinfo {year} {2011})}\BibitemShut
	{NoStop}%
	\bibitem [{\citenamefont {Langen}\ \emph {et~al.}(2016)\citenamefont {Langen},
		\citenamefont {Gasenzer},\ and\ \citenamefont {Schmiedmayer}}]{Langen2016}%
	\BibitemOpen
	\bibfield  {author} {\bibinfo {author} {\bibfnamefont {T.}~\bibnamefont
			{Langen}}, \bibinfo {author} {\bibfnamefont {T.}~\bibnamefont {Gasenzer}},\
		and\ \bibinfo {author} {\bibfnamefont {J.}~\bibnamefont {Schmiedmayer}},\
	}\bibfield  {title} {\bibinfo {title} {Prethermalization and universal
			dynamics in near-integrable quantum systems},\ }\href
	{https://doi.org/10.1088/1742-5468/2016/06/064009} {\bibfield  {journal}
		{\bibinfo  {journal} {J. Stat. Mech.}\ }\textbf {\bibinfo {volume} {2016}},\ \bibinfo {pages} {064009}
		(\bibinfo {year} {2016})}\BibitemShut {NoStop}%
  %       \bibitem [{\citenamefont {Ketterle}\ \emph {et~al.}(1999)\citenamefont
		% {Ketterle}, \citenamefont {Durfee},\ and\ \citenamefont
		% {Stamper-Kurn}}]{Ketterle1999}%
  %       \BibitemOpen
  %       \bibfield  {author} {\bibinfo {author} {\bibfnamefont {W.}~\bibnamefont
		% 	{Ketterle}}, \bibinfo {author} {\bibfnamefont {D.~S.}\ \bibnamefont
		% 	{Durfee}},\ and\ \bibinfo {author} {\bibfnamefont {D.~M.}\ \bibnamefont
		% 	{Stamper-Kurn}},\ } in\ 
  % \href {https://doi.org/10.3254/978-1-61499-225-7-67} {\emph {\bibinfo{booktitle} {Proceedings of the International School of Physics "Enrico Fermi"}}},\ Vol.\ \bibinfo {volume} {140}\ (\bibinfo  {publisher} {IOS Press},\ \bibinfo {year} {1999})\ pp.\ \bibinfo {pages} {67--164}\BibitemShut
  % {NoStop}%
	\bibitem [{\citenamefont {Rigol}\ \emph {et~al.}(2008)\citenamefont {Rigol},
		\citenamefont {Dunjko},\ and\ \citenamefont {Olshanii}}]{Rigol2008}%
	\BibitemOpen
	\bibfield  {author} {\bibinfo {author} {\bibfnamefont {M.}~\bibnamefont
			{Rigol}}, \bibinfo {author} {\bibfnamefont {V.}~\bibnamefont {Dunjko}},\ and\
		\bibinfo {author} {\bibfnamefont {M.}~\bibnamefont {Olshanii}},\ }\bibfield
	{title} {\bibinfo {title} {Thermalization and its mechanism for generic
			isolated quantum systems},\ }\href {https://doi.org/10.1038/nature06838}
	{\bibfield  {journal} {\bibinfo  {journal} {Nature (London)}\ }\textbf {\bibinfo
			{volume} {452}},\ \bibinfo {pages} {854} (\bibinfo {year}
		{2008})}\BibitemShut {NoStop}%
	\bibitem [{\citenamefont {Trotzky}\ \emph {et~al.}(2012)\citenamefont
		{Trotzky}, \citenamefont {Chen}, \citenamefont {Flesch}, \citenamefont
		{McCulloch}, \citenamefont {Schollw{\"o}ck}, \citenamefont {Eisert},\ and\
		\citenamefont {Bloch}}]{trotzky2012probing}%
	\BibitemOpen
	\bibfield  {author} {\bibinfo {author} {\bibfnamefont {S.}~\bibnamefont
			{Trotzky}}, \bibinfo {author} {\bibfnamefont {Y.-A.}\ \bibnamefont {Chen}},
		\bibinfo {author} {\bibfnamefont {A.}~\bibnamefont {Flesch}}, \bibinfo
		{author} {\bibfnamefont {I.~P.}\ \bibnamefont {McCulloch}}, \bibinfo {author}
		{\bibfnamefont {U.}~\bibnamefont {Schollw{\"o}ck}}, \bibinfo {author}
		{\bibfnamefont {J.}~\bibnamefont {Eisert}},\ and\ \bibinfo {author}
		{\bibfnamefont {I.}~\bibnamefont {Bloch}},\ }\bibfield  {title} {\bibinfo
		{title} {Probing the relaxation towards equilibrium in an isolated strongly
			correlated one-dimensional {B}ose gas},\ }\href
	{https://doi.org/https://doi.org/10.1038/nphys2232} {\bibfield  {journal}
		{\bibinfo  {journal} {Nat. Phys.}\ }\textbf {\bibinfo {volume} {8}},\
		\bibinfo {pages} {325} (\bibinfo {year} {2012})}\BibitemShut {NoStop}%
	\bibitem [{\citenamefont {Gring}\ \emph {et~al.}(2012)\citenamefont {Gring},
		\citenamefont {Kuhnert}, \citenamefont {Langen}, \citenamefont {Kitagawa},
		\citenamefont {Rauer}, \citenamefont {Schreitl}, \citenamefont {Mazets},
		\citenamefont {Smith}, \citenamefont {Demler},\ and\ \citenamefont
		{Schmiedmayer}}]{gring2012relaxation}%
	\BibitemOpen
	\bibfield  {author} {\bibinfo {author} {\bibfnamefont {M.}~\bibnamefont
			{Gring}}, \bibinfo {author} {\bibfnamefont {M.}~\bibnamefont {Kuhnert}},
		\bibinfo {author} {\bibfnamefont {T.}~\bibnamefont {Langen}}, \bibinfo
		{author} {\bibfnamefont {T.}~\bibnamefont {Kitagawa}}, \bibinfo {author}
		{\bibfnamefont {B.}~\bibnamefont {Rauer}}, \bibinfo {author} {\bibfnamefont
			{M.}~\bibnamefont {Schreitl}}, \bibinfo {author} {\bibfnamefont
			{I.}~\bibnamefont {Mazets}}, \bibinfo {author} {\bibfnamefont {D.~A.}\
			\bibnamefont {Smith}}, \bibinfo {author} {\bibfnamefont {E.}~\bibnamefont
			{Demler}},\ and\ \bibinfo {author} {\bibfnamefont {J.}~\bibnamefont
			{Schmiedmayer}},\ }\bibfield  {title} {\bibinfo {title} {Relaxation and
			prethermalization in an isolated quantum system},\ }\href
	{https://doi.org/10.1126/science.1224953} {\bibfield  {journal} {\bibinfo
			{journal} {Science}\ }\textbf {\bibinfo {volume} {337}},\ \bibinfo {pages}
		{1318} (\bibinfo {year} {2012})}\BibitemShut {NoStop}%
	\bibitem [{\citenamefont {Langen}\ \emph {et~al.}(2015)\citenamefont {Langen},
		\citenamefont {Geiger},\ and\ \citenamefont {Schmiedmayer}}]{Langen2015}%
	\BibitemOpen
	\bibfield  {author} {\bibinfo {author} {\bibfnamefont {T.}~\bibnamefont
			{Langen}}, \bibinfo {author} {\bibfnamefont {R.}~\bibnamefont {Geiger}},\
		and\ \bibinfo {author} {\bibfnamefont {J.}~\bibnamefont {Schmiedmayer}},\
	}\bibfield  {title} {\bibinfo {title} {Ultracold atoms out of equilibrium},\
	}\href {https://doi.org/10.1146/annurev-conmatphys-031214-014548} {\bibfield
		{journal} {\bibinfo  {journal} {Ann. Rev. Condens. Matter Phys.}\
		}\textbf {\bibinfo {volume} {6}},\ \bibinfo {pages} {201} (\bibinfo {year}
		{2015})}\BibitemShut {NoStop}%
	\bibitem [{\citenamefont {Seman}\ \emph {et~al.}(2011)\citenamefont {Seman},
		\citenamefont {Henn}, \citenamefont {Shiozaki}, \citenamefont {Roati},
		\citenamefont {Poveda-Cuevas}, \citenamefont {Magalh\"{a}es}, \citenamefont
		{Yukalov}, \citenamefont {Tsubota}, \citenamefont {Kobayashi}, \citenamefont
		{Kasamatsu},\ and\ \citenamefont {Bagnato}}]{Seman2011}%
	\BibitemOpen
	\bibfield  {author} {\bibinfo {author} {\bibfnamefont {J.}~\bibnamefont
			{Seman}}, \bibinfo {author} {\bibfnamefont {E.}~\bibnamefont {Henn}},
		\bibinfo {author} {\bibfnamefont {R.}~\bibnamefont {Shiozaki}}, \bibinfo
		{author} {\bibfnamefont {G.}~\bibnamefont {Roati}}, \bibinfo {author}
		{\bibfnamefont {F.}~\bibnamefont {Poveda-Cuevas}}, \bibinfo {author}
		{\bibfnamefont {K.}~\bibnamefont {Magalh\"{a}es}}, \bibinfo {author}
		{\bibfnamefont {V.}~\bibnamefont {Yukalov}}, \bibinfo {author} {\bibfnamefont
			{M.}~\bibnamefont {Tsubota}}, \bibinfo {author} {\bibfnamefont
			{M.}~\bibnamefont {Kobayashi}}, \bibinfo {author} {\bibfnamefont
			{K.}~\bibnamefont {Kasamatsu}},\ and\ \bibinfo {author} {\bibfnamefont
			{V.}~\bibnamefont {Bagnato}},\ }\bibfield  {title} {\bibinfo {title} {Route
			to turbulence in a trapped {B}ose-{E}instein condensate},\ }\href
	{https://dx.doi.org/10.1002/lapl.201110052} {\bibfield  {journal} {\bibinfo
			{journal} {Laser Phys. Lett.}\ }\textbf {\bibinfo {volume} {8}},\ \bibinfo
		{pages} {691} (\bibinfo {year} {2011})}\BibitemShut {NoStop}%
	\bibitem [{\citenamefont {Madeira}\ \emph {et~al.}(2020)\citenamefont
		{Madeira}, \citenamefont {Caracanhas}, \citenamefont {dos Santos},\ and\
		\citenamefont {Bagnato}}]{Madeira2020}%
	\BibitemOpen
	\bibfield  {author} {\bibinfo {author} {\bibfnamefont {L.}~\bibnamefont
			{Madeira}}, \bibinfo {author} {\bibfnamefont {M.}~\bibnamefont {Caracanhas}},
		\bibinfo {author} {\bibfnamefont {F.}~\bibnamefont {dos Santos}},\ and\
		\bibinfo {author} {\bibfnamefont {V.}~\bibnamefont {Bagnato}},\ }\bibfield
	{title} {\bibinfo {title} {Quantum turbulence in quantum gases},\ }\href
	{https://doi.org/10.1146/annurev-conmatphys-031119-050821} {\bibfield
		{journal} {\bibinfo  {journal} {Ann. Rev. Condens. Matter Phys.}\
		}\textbf {\bibinfo {volume} {11}},\ \bibinfo {pages} {37} (\bibinfo {year}
		{2020})}\BibitemShut {NoStop}%
	\bibitem [{\citenamefont {Henn}\ \emph {et~al.}(2009)\citenamefont {Henn},
		\citenamefont {Seman}, \citenamefont {Roati}, \citenamefont {Magalh\~aes},\
		and\ \citenamefont {Bagnato}}]{PhysRevLett.103.045301}%
	\BibitemOpen
	\bibfield  {author} {\bibinfo {author} {\bibfnamefont {E.~A.~L.}\
			\bibnamefont {Henn}}, \bibinfo {author} {\bibfnamefont {J.~A.}\ \bibnamefont
			{Seman}}, \bibinfo {author} {\bibfnamefont {G.}~\bibnamefont {Roati}},
		\bibinfo {author} {\bibfnamefont {K.~M.~F.}\ \bibnamefont {Magalh\~aes}},\
		and\ \bibinfo {author} {\bibfnamefont {V.~S.}\ \bibnamefont {Bagnato}},\
	}\bibfield  {title} {\bibinfo {title} {Emergence of turbulence in an
			oscillating {B}ose-{E}instein condensate},\ }\href
	{https://doi.org/10.1103/PhysRevLett.103.045301} {\bibfield  {journal}
		{\bibinfo  {journal} {Phys. Rev. Lett.}\ }\textbf {\bibinfo {volume} {103}},\
		\bibinfo {pages} {045301} (\bibinfo {year} {2009})}\BibitemShut {NoStop}%
	\bibitem [{\citenamefont {Thompson}\ \emph {et~al.}(2013)\citenamefont
		{Thompson}, \citenamefont {Bagnato}, \citenamefont {Telles}, \citenamefont
		{Caracanhas}, \citenamefont {dos Santos},\ and\ \citenamefont
		{Bagnato}}]{Thompson_2014}%
	\BibitemOpen
	\bibfield  {author} {\bibinfo {author} {\bibfnamefont {K.~J.}\ \bibnamefont
			{Thompson}}, \bibinfo {author} {\bibfnamefont {G.~G.}\ \bibnamefont
			{Bagnato}}, \bibinfo {author} {\bibfnamefont {G.~D.}\ \bibnamefont {Telles}},
		\bibinfo {author} {\bibfnamefont {M.~A.}\ \bibnamefont {Caracanhas}},
		\bibinfo {author} {\bibfnamefont {F.~E.~A.}\ \bibnamefont {dos Santos}},\
		and\ \bibinfo {author} {\bibfnamefont {V.~S.}\ \bibnamefont {Bagnato}},\
	}\bibfield  {title} {\bibinfo {title} {Evidence of power law behavior in the
			momentum distribution of a turbulent trapped {B}ose–{E}instein
			condensate},\ }\href {https://doi.org/10.1088/1612-2011/11/1/015501}
	{\bibfield  {journal} {\bibinfo  {journal} {Laser Phys. Lett.}\ }\textbf
		{\bibinfo {volume} {11}},\ \bibinfo {pages} {015501} (\bibinfo {year}
		{2013})}\BibitemShut {NoStop}%
	\bibitem [{\citenamefont {Navon}\ \emph {et~al.}(2016)\citenamefont {Navon},
		\citenamefont {Gaunt}, \citenamefont {Smith},\ and\ \citenamefont
		{Hadzibabic}}]{navon2016emergence}%
	\BibitemOpen
	\bibfield  {author} {\bibinfo {author} {\bibfnamefont {N.}~\bibnamefont
			{Navon}}, \bibinfo {author} {\bibfnamefont {A.~L.}\ \bibnamefont {Gaunt}},
		\bibinfo {author} {\bibfnamefont {R.~P.}\ \bibnamefont {Smith}},\ and\
		\bibinfo {author} {\bibfnamefont {Z.}~\bibnamefont {Hadzibabic}},\ }\bibfield
	{title} {\bibinfo {title} {Emergence of a turbulent cascade in a quantum
			gas},\ }\href {https://doi.org/https://doi.org/10.1038/nature20114}
	{\bibfield  {journal} {\bibinfo  {journal} {Nature (London)}\ }\textbf {\bibinfo
			{volume} {539}},\ \bibinfo {pages} {72} (\bibinfo {year} {2016})}\BibitemShut
	{NoStop}%
	\bibitem [{\citenamefont {Nowak}\ \emph {et~al.}(2012)\citenamefont {Nowak},
		\citenamefont {Schole}, \citenamefont {Sexty},\ and\ \citenamefont
		{Gasenzer}}]{Nowak2012}%
	\BibitemOpen
	\bibfield  {author} {\bibinfo {author} {\bibfnamefont {B.}~\bibnamefont
			{Nowak}}, \bibinfo {author} {\bibfnamefont {J.}~\bibnamefont {Schole}},
		\bibinfo {author} {\bibfnamefont {D.}~\bibnamefont {Sexty}},\ and\ \bibinfo
		{author} {\bibfnamefont {T.}~\bibnamefont {Gasenzer}},\ }\bibfield  {title}
	{\bibinfo {title} {Nonthermal fixed points, vortex statistics, and superfluid
			turbulence in an ultracold {B}ose gas},\ }\href
	{https://doi.org/10.1103/PhysRevA.85.043627} {\bibfield  {journal} {\bibinfo
			{journal} {Phys. Rev. A}\ }\textbf {\bibinfo {volume} {85}},\ \bibinfo
		{pages} {043627} (\bibinfo {year} {2012})}\BibitemShut {NoStop}%
	\bibitem [{\citenamefont {Kwon}\ \emph {et~al.}(2014)\citenamefont {Kwon},
		\citenamefont {Moon}, \citenamefont {Choi}, \citenamefont {Seo},\ and\
		\citenamefont {Shin}}]{kwon2014relaxation}%
	\BibitemOpen
	\bibfield  {author} {\bibinfo {author} {\bibfnamefont {W.~J.}\ \bibnamefont
			{Kwon}}, \bibinfo {author} {\bibfnamefont {G.}~\bibnamefont {Moon}}, \bibinfo
		{author} {\bibfnamefont {J.~Y.}\ \bibnamefont {Choi}}, \bibinfo {author}
		{\bibfnamefont {S.~W.}\ \bibnamefont {Seo}},\ and\ \bibinfo {author}
		{\bibfnamefont {Y.~I.}\ \bibnamefont {Shin}},\ }\bibfield  {title} {\bibinfo
		{title} {Relaxation of superfluid turbulence in highly oblate
			{B}ose-{E}instein condensates},\ }\href
	{https://doi.org/10.1103/PhysRevA.90.063627} {\bibfield  {journal} {\bibinfo
			{journal} {Phys. Rev. A}\ }\textbf {\bibinfo {volume} {90}},\ \bibinfo
		{pages} {063627} (\bibinfo {year} {2014})}\BibitemShut {NoStop}%
	\bibitem [{\citenamefont {Nowak}\ \emph {et~al.}(2014)\citenamefont {Nowak},
		\citenamefont {Schole},\ and\ \citenamefont {Gasenzer}}]{Nowak2014}%
	\BibitemOpen
	\bibfield  {author} {\bibinfo {author} {\bibfnamefont {B.}~\bibnamefont
			{Nowak}}, \bibinfo {author} {\bibfnamefont {J.}~\bibnamefont {Schole}},\ and\
		\bibinfo {author} {\bibfnamefont {T.}~\bibnamefont {Gasenzer}},\ }\bibfield
	{title} {\bibinfo {title} {Universal dynamics on the way to thermalization},\
	}\href {https://doi.org/10.1088/1367-2630/16/9/093052} {\bibfield  {journal}
		{\bibinfo  {journal} {New J. Phys.}\ }\textbf {\bibinfo {volume}
			{16}},\ \bibinfo {pages} {093052} (\bibinfo {year} {2014})}\BibitemShut
	{NoStop}%
	\bibitem [{\citenamefont {Reeves}\ \emph {et~al.}(2022)\citenamefont {Reeves},
		\citenamefont {Goddard-Lee}, \citenamefont {Gauthier}, \citenamefont
		{Stockdale}, \citenamefont {Salman}, \citenamefont {Edmonds}, \citenamefont
		{Yu}, \citenamefont {Bradley}, \citenamefont {Baker}, \citenamefont
		{Rubinsztein-Dunlop}, \citenamefont {Davis},\ and\ \citenamefont
		{Neely}}]{Reeves2022}%
	\BibitemOpen
	\bibfield  {author} {\bibinfo {author} {\bibfnamefont {M.~T.}\ \bibnamefont
			{Reeves}}, \bibinfo {author} {\bibfnamefont {K.}~\bibnamefont {Goddard-Lee}},
		\bibinfo {author} {\bibfnamefont {G.}~\bibnamefont {Gauthier}}, \bibinfo
		{author} {\bibfnamefont {O.~R.}\ \bibnamefont {Stockdale}}, \bibinfo {author}
		{\bibfnamefont {H.}~\bibnamefont {Salman}}, \bibinfo {author} {\bibfnamefont
			{T.}~\bibnamefont {Edmonds}}, \bibinfo {author} {\bibfnamefont
			{X.}~\bibnamefont {Yu}}, \bibinfo {author} {\bibfnamefont {A.~S.}\
			\bibnamefont {Bradley}}, \bibinfo {author} {\bibfnamefont {M.}~\bibnamefont
			{Baker}}, \bibinfo {author} {\bibfnamefont {H.}~\bibnamefont
			{Rubinsztein-Dunlop}}, \bibinfo {author} {\bibfnamefont {M.~J.}\ \bibnamefont
			{Davis}},\ and\ \bibinfo {author} {\bibfnamefont {T.~W.}\ \bibnamefont
			{Neely}},\ }\bibfield  {title} {\bibinfo {title} {Turbulent relaxation to
			equilibrium in a two-dimensional quantum vortex gas},\ }\href
	{https://doi.org/10.1103/PhysRevX.12.011031} {\bibfield  {journal} {\bibinfo
			{journal} {Phys. Rev. X}\ }\textbf {\bibinfo {volume} {12}},\ \bibinfo
		{pages} {011031} (\bibinfo {year} {2022})}\BibitemShut {NoStop}%
	\bibitem [{\citenamefont {Vicsek}\ and\ \citenamefont
		{Family}(1984)}]{Famili_1}%
	\BibitemOpen
	\bibfield  {author} {\bibinfo {author} {\bibfnamefont {T.}~\bibnamefont
			{Vicsek}}\ and\ \bibinfo {author} {\bibfnamefont {F.}~\bibnamefont
			{Family}},\ }\bibfield  {title} {\bibinfo {title} {Dynamic scaling for
			aggregation of clusters},\ }\href
	{https://doi.org/10.1103/PhysRevLett.52.1669} {\bibfield  {journal} {\bibinfo
			{journal} {Phys. Rev. Lett.}\ }\textbf {\bibinfo {volume} {52}},\ \bibinfo
		{pages} {1669} (\bibinfo {year} {1984})}\BibitemShut {NoStop}%
	\bibitem [{\citenamefont {Family}\ and\ \citenamefont
		{Vicsek}(1985)}]{Famili_2}%
	\BibitemOpen
	\bibfield  {author} {\bibinfo {author} {\bibfnamefont {F.}~\bibnamefont
			{Family}}\ and\ \bibinfo {author} {\bibfnamefont {T.}~\bibnamefont
			{Vicsek}},\ }\bibfield  {title} {\bibinfo {title} {Scaling of the active zone
			in the {E}den process on percolation networks and the ballistic deposition
			model},\ }\href {https://doi.org/10.1088/0305-4470/18/2/005} {\bibfield
		{journal} {\bibinfo  {journal} {J. Phys. A}\ }\textbf {\bibinfo {volume}
			{18}},\ \bibinfo {pages} {L75} (\bibinfo {year} {1985})}\BibitemShut
	{NoStop}%
	\bibitem [{\citenamefont {Orioli}\ \emph {et~al.}(2015)\citenamefont {Orioli},
		\citenamefont {Boguslavski},\ and\ \citenamefont {Berges}}]{Orioli2015}%
	\BibitemOpen
	\bibfield  {author} {\bibinfo {author} {\bibfnamefont {A.~P.}\ \bibnamefont
			{Pineiro Orioli}}, \bibinfo {author} {\bibfnamefont {K.}~\bibnamefont
			{Boguslavski}},\ and\ \bibinfo {author} {\bibfnamefont {J.}~\bibnamefont
			{Berges}},\ }\bibfield  {title} {\bibinfo {title} {Universal self-similar
			dynamics of relativistic and nonrelativistic field theories near nonthermal
			fixed points},\ }\href {https://doi.org/10.1103/PhysRevD.92.025041}
	{\bibfield  {journal} {\bibinfo  {journal} {Phys. Rev. D}\ }\textbf
		{\bibinfo {volume} {92}},\ \bibinfo {pages} {025041} (\bibinfo {year}
		{2015})}\BibitemShut {NoStop}%
	\bibitem [{\citenamefont {Schmied}\ \emph {et~al.}(2019)\citenamefont
		{Schmied}, \citenamefont {Mikheev},\ and\ \citenamefont
		{Gasenzer}}]{Schmied2019}%
	\BibitemOpen
	\bibfield  {author} {\bibinfo {author} {\bibfnamefont {C.-M.}\ \bibnamefont
			{Schmied}}, \bibinfo {author} {\bibfnamefont {A.~N.}\ \bibnamefont
			{Mikheev}},\ and\ \bibinfo {author} {\bibfnamefont {T.}~\bibnamefont
			{Gasenzer}},\ }\bibfield  {title} {\bibinfo {title} {Non-thermal fixed
			points: Universal dynamics far from equilibrium},\ }\href
	{https://doi.org/10.1142/S0217751X19410069} {\bibfield  {journal} {\bibinfo
			{journal} {Int. J. Mod. Phys. A}\ }\textbf {\bibinfo
			{volume} {34}},\ \bibinfo {pages} {1941006} (\bibinfo {year}
		{2019})}\BibitemShut {NoStop}%
	\bibitem [{\citenamefont {Mikheev}\ \emph {et~al.}(2023)\citenamefont
		{Mikheev}, \citenamefont {Siovitz},\ and\ \citenamefont
		{Gasenzer}}]{Mikheev2023}%
	\BibitemOpen
	\bibfield  {author} {\bibinfo {author} {\bibfnamefont {A.~N.}\ \bibnamefont
			{Mikheev}}, \bibinfo {author} {\bibfnamefont {I.}~\bibnamefont {Siovitz}},\
		and\ \bibinfo {author} {\bibfnamefont {T.}~\bibnamefont {Gasenzer}},\
	}\bibfield  {title} {\bibinfo {title} {Universal dynamics and non-thermal
			fixed points in quantum fluids far from equilibrium},\ }\href
	{https://doi.org/10.1140/epjs/s11734-023-00974-7} {\bibfield  {journal}
		{\bibinfo  {journal} {Eur. Phys. J. Special Topics}\ }\textbf
		{\bibinfo {volume} {232}},\ \bibinfo {pages} {3393} (\bibinfo {year}
		{2023})}\BibitemShut {NoStop}%
	\bibitem [{\citenamefont {Nazarenko}(2011)}]{nazarenko2011wave}%
	\BibitemOpen
	\bibfield  {author} {\bibinfo {author} {\bibfnamefont {S.}~\bibnamefont
			{Nazarenko}},\ }\href {https://books.google.com.br/books?id=cAHZ2aEMqgYC}
	{\emph {\bibinfo {title} {Wave Turbulence}}},\ Lecture notes in physics\
	(\bibinfo  {publisher} {Springer, Berlin, Heidelberg},\ \bibinfo {year}
	{2011})\BibitemShut {NoStop}%
	\bibitem [{\citenamefont {Zhu}\ \emph {et~al.}(2023{\natexlab{a}})\citenamefont
		{Zhu}, \citenamefont {Semisalov}, \citenamefont {Krstulovic},\ and\
		\citenamefont {Nazarenko}}]{Zhu2023}%
	\BibitemOpen
	\bibfield  {author} {\bibinfo {author} {\bibfnamefont {Y.}~\bibnamefont
			{Zhu}}, \bibinfo {author} {\bibfnamefont {B.}~\bibnamefont {Semisalov}},
		\bibinfo {author} {\bibfnamefont {G.}~\bibnamefont {Krstulovic}},\ and\
		\bibinfo {author} {\bibfnamefont {S.}~\bibnamefont {Nazarenko}},\ }\bibfield
	{title} {\bibinfo {title} {Direct and inverse cascades in turbulent
			{B}ose-{E}instein condensates},\ }\href
	{https://doi.org/10.1103/PhysRevLett.130.133001} {\bibfield  {journal}
		{\bibinfo  {journal} {Phys. Rev. Lett.}\ }\textbf {\bibinfo {volume}
			{130}},\ \bibinfo {pages} {133001} (\bibinfo {year}
		{2023}{\natexlab{a}})}\BibitemShut {NoStop}%
	\bibitem [{\citenamefont {Zhu}\ \emph {et~al.}(2023{\natexlab{b}})\citenamefont
		{Zhu}, \citenamefont {Semisalov}, \citenamefont {Krstulovic},\ and\
		\citenamefont {Nazarenko}}]{Zhu2023PRE}%
	\BibitemOpen
	\bibfield  {author} {\bibinfo {author} {\bibfnamefont {Y.}~\bibnamefont
			{Zhu}}, \bibinfo {author} {\bibfnamefont {B.}~\bibnamefont {Semisalov}},
		\bibinfo {author} {\bibfnamefont {G.}~\bibnamefont {Krstulovic}},\ and\
		\bibinfo {author} {\bibfnamefont {S.}~\bibnamefont {Nazarenko}},\ }\bibfield
	{title} {\bibinfo {title} {Self-similar evolution of wave turbulence in
			{G}ross-{P}itaevskii system},\ }\href
	{https://doi.org/10.1103/PhysRevE.108.064207} {\bibfield  {journal} {\bibinfo
			{journal} {Phys. Rev. E}\ }\textbf {\bibinfo {volume} {108}},\ \bibinfo
		{pages} {064207} (\bibinfo {year} {2023}{\natexlab{b}})}\BibitemShut
	{NoStop}%
	\bibitem [{\citenamefont {Garc\'{\i}a-Orozco}\ \emph
		{et~al.}(2022)\citenamefont {Garc\'{\i}a-Orozco}, \citenamefont {Madeira},
		\citenamefont {Moreno-Armijos}, \citenamefont {Fritsch}, \citenamefont
		{Tavares}, \citenamefont {Castilho}, \citenamefont {Cidrim}, \citenamefont
		{Roati},\ and\ \citenamefont {Bagnato}}]{PhysRevA2022}%
	\BibitemOpen
	\bibfield  {author} {\bibinfo {author} {\bibfnamefont {A.~D.}\ \bibnamefont
			{Garc\'{\i}a-Orozco}}, \bibinfo {author} {\bibfnamefont {L.}~\bibnamefont
			{Madeira}}, \bibinfo {author} {\bibfnamefont {M.~A.}\ \bibnamefont
			{Moreno-Armijos}}, \bibinfo {author} {\bibfnamefont {A.~R.}\ \bibnamefont
			{Fritsch}}, \bibinfo {author} {\bibfnamefont {P.~E.~S.}\ \bibnamefont
			{Tavares}}, \bibinfo {author} {\bibfnamefont {P.~C.~M.}\ \bibnamefont
			{Castilho}}, \bibinfo {author} {\bibfnamefont {A.}~\bibnamefont {Cidrim}},
		\bibinfo {author} {\bibfnamefont {G.}~\bibnamefont {Roati}},\ and\ \bibinfo
		{author} {\bibfnamefont {V.~S.}\ \bibnamefont {Bagnato}},\ }\bibfield
	{title} {\bibinfo {title} {Universal dynamics of a turbulent superfluid
			{B}ose gas},\ }\href {https://doi.org/10.1103/PhysRevA.106.023314} {\bibfield
		{journal} {\bibinfo  {journal} {Phys. Rev. A}\ }\textbf {\bibinfo {volume}
			{106}},\ \bibinfo {pages} {023314} (\bibinfo {year} {2022})}\BibitemShut
	{NoStop}%
	\bibitem [{Sup()}]{SupMat}%
	\BibitemOpen
	\href@noop {} {}\bibinfo {note} {See Supplemental Material for more details on the experimental method, measurement of the momentum distributions, determination of the universal
		exponents, and particle cascades.}\BibitemShut {Stop}%
	\bibitem [{\citenamefont {Madeira}\ \emph {et~al.}(2023)\citenamefont
		{Madeira}, \citenamefont {Garc{\'\i}a-Orozco}, \citenamefont
		{Moreno-Armijos}, \citenamefont {Fritsch},\ and\ \citenamefont
		{Bagnato}}]{madeira2023differential}%
	\BibitemOpen
	\bibfield  {author} {\bibinfo {author} {\bibfnamefont {L.}~\bibnamefont
			{Madeira}}, \bibinfo {author} {\bibfnamefont {A.}~\bibnamefont
			{Garc{\'\i}a-Orozco}}, \bibinfo {author} {\bibfnamefont {M.}~\bibnamefont
			{Moreno-Armijos}}, \bibinfo {author} {\bibfnamefont {A.}~\bibnamefont
			{Fritsch}},\ and\ \bibinfo {author} {\bibfnamefont {V.}~\bibnamefont
			{Bagnato}},\ }\bibfield {title} {\bibinfo {title} {Universal scaling in far-from-equilibrium quantum systems: An equivalent differential approach},\ }\href
	{https://www.pnas.org/doi/10.1073/pnas.2404828121} {\bibfield  {journal}
		{\bibinfo  {journal} {Proc. Natl. Acad. Sci. U.S.A.}\ }\textbf {\bibinfo {volume} {121}},\
		\bibinfo {pages} {e2404828121} (\bibinfo {year} {2024})}\BibitemShut {NoStop}%
    \bibitem [{Sup()}]{Temperature}%
	\BibitemOpen
	\href@noop {} {}\bibinfo {note} {A simplified temperature analysis in the time window $t > t_4$ indicates that the system reached a temperature $T = 135(5)$ nK, while the critical temperature is $T_c \approx 130$ nK. This suggests that the system did not reach thermal equilibrium during repopulation. In the final stage ($t > t_4$), the cloud would still be on its path to thermalization above $T_c$. In the case of larger excitation amplitudes, the repopulation peak is hindered because the system reaches the critical temperature even before repopulation. It is worth noting that all stages are of the order of $100$ ms: $t_1-t_0=70$ ms, $t_2 - t_1 = 90$ ms, $t_3 - t_2 = 100$ ms, and $t_4 - t_3 = 110$ ms, which contributes to our knowledge of the characteristic times of equilibration in this system.}\BibitemShut {Stop}%
	\bibitem [{\citenamefont {Barenghi}\ \emph {et~al.}(2023)\citenamefont
		{Barenghi}, \citenamefont {Middleton-Spencer}, \citenamefont {Galantucci},\
		and\ \citenamefont {Parker}}]{barenghi2023types}%
	\BibitemOpen
	\bibfield  {author} {\bibinfo {author} {\bibfnamefont {C.~F.}\ \bibnamefont
			{Barenghi}}, \bibinfo {author} {\bibfnamefont {H.}~\bibnamefont
			{Middleton-Spencer}}, \bibinfo {author} {\bibfnamefont {L.}~\bibnamefont
			{Galantucci}},\ and\ \bibinfo {author} {\bibfnamefont {N.}~\bibnamefont
			{Parker}},\ }\bibfield  {title} {\bibinfo {title} {Types of quantum
			turbulence},\ } \href{https://doi.org/10.1116/5.0146107} {\bibfield  {journal} {\bibinfo  {journal} {AVS Quantum Sci.}\ }\textbf {\bibinfo {volume} {5}},\ \bibinfo {pages} {025601}\ (\bibinfo {year} {2023})}\BibitemShut {NoStop}
	\bibitem [{\citenamefont {Nicklas}\ \emph {et~al.}(2015)\citenamefont
		{Nicklas}, \citenamefont {Karl}, \citenamefont {H\"ofer}, \citenamefont
		{Johnson}, \citenamefont {Muessel}, \citenamefont {Strobel}, \citenamefont
		{Tomkovi\ifmmode~\check{c}\else \v{c}\fi{}}, \citenamefont {Gasenzer},\ and\
		\citenamefont {Oberthaler}}]{Nicklas2015observationscaling}%
	\BibitemOpen
	\bibfield  {author} {\bibinfo {author} {\bibfnamefont {E.}~\bibnamefont
			{Nicklas}}, \bibinfo {author} {\bibfnamefont {M.}~\bibnamefont {Karl}},
		\bibinfo {author} {\bibfnamefont {M.}~\bibnamefont {H\"ofer}}, \bibinfo
		{author} {\bibfnamefont {A.}~\bibnamefont {Johnson}}, \bibinfo {author}
		{\bibfnamefont {W.}~\bibnamefont {Muessel}}, \bibinfo {author} {\bibfnamefont
			{H.}~\bibnamefont {Strobel}}, \bibinfo {author} {\bibfnamefont
			{J.}~\bibnamefont {Tomkovi\ifmmode~\check{c}\else \v{c}\fi{}}}, \bibinfo
		{author} {\bibfnamefont {T.}~\bibnamefont {Gasenzer}},\ and\ \bibinfo
		{author} {\bibfnamefont {M.~K.}\ \bibnamefont {Oberthaler}},\ }\bibfield
	{title} {\bibinfo {title} {Observation of scaling in the dynamics of a
			strongly quenched quantum gas},\ }\href
	{https://doi.org/10.1103/PhysRevLett.115.245301} {\bibfield  {journal}
		{\bibinfo  {journal} {Phys. Rev. Lett.}\ }\textbf {\bibinfo {volume} {115}},\
		\bibinfo {pages} {245301} (\bibinfo {year} {2015})}\BibitemShut {NoStop}%
	\bibitem [{\citenamefont {Pr{\"u}fer}\ \emph {et~al.}(2018)\citenamefont
		{Pr{\"u}fer}, \citenamefont {Kunkel}, \citenamefont {Strobel}, \citenamefont
		{Lannig}, \citenamefont {Linnemann}, \citenamefont {Schmied}, \citenamefont
		{Berges}, \citenamefont {Gasenzer},\ and\ \citenamefont
		{Oberthaler}}]{prufer2018observation}%
	\BibitemOpen
	\bibfield  {author} {\bibinfo {author} {\bibfnamefont {M.}~\bibnamefont
			{Pr{\"u}fer}}, \bibinfo {author} {\bibfnamefont {P.}~\bibnamefont {Kunkel}},
		\bibinfo {author} {\bibfnamefont {H.}~\bibnamefont {Strobel}}, \bibinfo
		{author} {\bibfnamefont {S.}~\bibnamefont {Lannig}}, \bibinfo {author}
		{\bibfnamefont {D.}~\bibnamefont {Linnemann}}, \bibinfo {author}
		{\bibfnamefont {C.-M.}\ \bibnamefont {Schmied}}, \bibinfo {author}
		{\bibfnamefont {J.}~\bibnamefont {Berges}}, \bibinfo {author} {\bibfnamefont
			{T.}~\bibnamefont {Gasenzer}},\ and\ \bibinfo {author} {\bibfnamefont
			{M.~K.}\ \bibnamefont {Oberthaler}},\ }\bibfield  {title} {\bibinfo {title}
		{Observation of universal dynamics in a spinor {B}ose gas far from
			equilibrium},\ }\href
	{https://doi.org/https://doi.org/10.1038/s41586-018-0659-0} {\bibfield
		{journal} {\bibinfo  {journal} {Nature (London)}\ }\textbf {\bibinfo {volume} {563}},\
		\bibinfo {pages} {217} (\bibinfo {year} {2018})}\BibitemShut {NoStop}%
	\bibitem [{\citenamefont {Erne}\ \emph {et~al.}(2018)\citenamefont {Erne},
		\citenamefont {B{\"u}cker}, \citenamefont {Gasenzer}, \citenamefont
		{Berges},\ and\ \citenamefont {Schmiedmayer}}]{erne2018universal}%
	\BibitemOpen
	\bibfield  {author} {\bibinfo {author} {\bibfnamefont {S.}~\bibnamefont
			{Erne}}, \bibinfo {author} {\bibfnamefont {R.}~\bibnamefont {B{\"u}cker}},
		\bibinfo {author} {\bibfnamefont {T.}~\bibnamefont {Gasenzer}}, \bibinfo
		{author} {\bibfnamefont {J.}~\bibnamefont {Berges}},\ and\ \bibinfo {author}
		{\bibfnamefont {J.}~\bibnamefont {Schmiedmayer}},\ }\bibfield  {title}
	{\bibinfo {title} {Universal dynamics in an isolated one-dimensional {B}ose
			gas far from equilibrium},\ }\href
	{https://doi.org/https://doi.org/10.1038/s41586-018-0667-0} {\bibfield
		{journal} {\bibinfo  {journal} {Nature (London)}\ }\textbf {\bibinfo {volume} {563}},\
		\bibinfo {pages} {225} (\bibinfo {year} {2018})}\BibitemShut {NoStop}%
	\bibitem [{\citenamefont {Eigen}\ \emph {et~al.}(2018)\citenamefont {Eigen},
		\citenamefont {Glidden}, \citenamefont {Lopes}, \citenamefont {Cornell},
		\citenamefont {Smith},\ and\ \citenamefont
		{Hadzibabic}}]{eigen2018universal}%
	\BibitemOpen
	\bibfield  {author} {\bibinfo {author} {\bibfnamefont {C.}~\bibnamefont
			{Eigen}}, \bibinfo {author} {\bibfnamefont {J.~A.}\ \bibnamefont {Glidden}},
		\bibinfo {author} {\bibfnamefont {R.}~\bibnamefont {Lopes}}, \bibinfo
		{author} {\bibfnamefont {E.~A.}\ \bibnamefont {Cornell}}, \bibinfo {author}
		{\bibfnamefont {R.~P.}\ \bibnamefont {Smith}},\ and\ \bibinfo {author}
		{\bibfnamefont {Z.}~\bibnamefont {Hadzibabic}},\ }\bibfield  {title}
	{\bibinfo {title} {Universal prethermal dynamics of {B}ose gases quenched to
			unitarity},\ }\href
	{https://doi.org/https://doi.org/10.1038/s41586-018-0674-1} {\bibfield
		{journal} {\bibinfo  {journal} {Nature (London)}\ }\textbf {\bibinfo {volume} {563}},\
		\bibinfo {pages} {221} (\bibinfo {year} {2018})}\BibitemShut {NoStop}%
	\bibitem [{\citenamefont {Glidden}\ \emph {et~al.}(2021)\citenamefont
		{Glidden}, \citenamefont {Eigen}, \citenamefont {Dogra}, \citenamefont
		{Hilker}, \citenamefont {Smith},\ and\ \citenamefont
		{Hadzibabic}}]{glidden2021bidirectional}%
	\BibitemOpen
	\bibfield  {author} {\bibinfo {author} {\bibfnamefont {J.~A.}\ \bibnamefont
			{Glidden}}, \bibinfo {author} {\bibfnamefont {C.}~\bibnamefont {Eigen}},
		\bibinfo {author} {\bibfnamefont {L.~H.}\ \bibnamefont {Dogra}}, \bibinfo
		{author} {\bibfnamefont {T.~A.}\ \bibnamefont {Hilker}}, \bibinfo {author}
		{\bibfnamefont {R.~P.}\ \bibnamefont {Smith}},\ and\ \bibinfo {author}
		{\bibfnamefont {Z.}~\bibnamefont {Hadzibabic}},\ }\bibfield  {title}
	{\bibinfo {title} {Bidirectional dynamic scaling in an isolated {B}ose gas
			far from equilibrium},\ }\href
	{https://doi.org/https://doi.org/10.1038/s41567-020-01114-x} {\bibfield
		{journal} {\bibinfo  {journal} {Nat. Phys.}\ }\textbf {\bibinfo {volume}
			{17}},\ \bibinfo {pages} {457} (\bibinfo {year} {2021})}\BibitemShut
	{NoStop}%
	\bibitem [{\citenamefont {Lannig}\ \emph {et~al.}(2023)\citenamefont {Lannig},
		\citenamefont {Pr{\"u}fer}, \citenamefont {Deller}, \citenamefont {Siovitz},
		\citenamefont {Dreher}, \citenamefont {Gasenzer}, \citenamefont {Strobel},\
		and\ \citenamefont {Oberthaler}}]{lannig2023observation}%
	\BibitemOpen
	\bibfield  {author} {\bibinfo {author} {\bibfnamefont {S.}~\bibnamefont
			{Lannig}}, \bibinfo {author} {\bibfnamefont {M.}~\bibnamefont {Pr{\"u}fer}},
		\bibinfo {author} {\bibfnamefont {Y.}~\bibnamefont {Deller}}, \bibinfo
		{author} {\bibfnamefont {I.}~\bibnamefont {Siovitz}}, \bibinfo {author}
		{\bibfnamefont {J.}~\bibnamefont {Dreher}}, \bibinfo {author} {\bibfnamefont
			{T.}~\bibnamefont {Gasenzer}}, \bibinfo {author} {\bibfnamefont
			{H.}~\bibnamefont {Strobel}},\ and\ \bibinfo {author} {\bibfnamefont {M.~K.}\
			\bibnamefont {Oberthaler}},\ }\bibfield  {title} {\bibinfo {title}
		{Observation of two non-thermal fixed points for the same microscopic
			symmetry},\ }\bibfield  {journal} {\bibinfo  {journal} {arXiv:2306.16497}\ }\href
	{https://doi.org/https://doi.org/10.48550/arXiv.2306.16497}
	{https://doi.org/10.48550/arXiv.2306.16497} (\bibinfo {year}
	{2023})\BibitemShut {NoStop}%
 \bibitem [{\citenamefont {Ga\l{}ka}\ \emph {et~al.}(2023)\citenamefont {Ga\l{}ka},
		\citenamefont {Christodoulou}, \citenamefont {Gazo}, \citenamefont {Karailiev},
		\citenamefont {Dogra}, \citenamefont {Schmitt},\
		and\ \citenamefont {Hadzibabic}}]{galka2022emergence}%
	\BibitemOpen
	\bibfield {author} 
        {\bibinfo {author} {\bibfnamefont {M.}~\bibnamefont{Ga\l{}ka}}, 
        \bibinfo {author} {\bibfnamefont {P.}~\bibnamefont {Christodoulou}},
	\bibinfo {author} {\bibfnamefont {M.}~\bibnamefont {Gazo}}, 
        \bibinfo {author} {\bibfnamefont {A.}~\bibnamefont {Karailiev}}, 
        \bibinfo {author} {\bibfnamefont {N.}~\bibnamefont {Dogra}}, 
        \bibinfo {author} {\bibfnamefont {J.}~\bibnamefont {Schmitt}}, 
        \ and\ \bibinfo {author} {\bibfnamefont {Z.}\ \bibnamefont {Hadzibabic}},\ }
        \bibfield  {title}
	{\bibinfo {title} {Emergence of isotropy and dynamic scaling in 2{D} wave turbulence in a homogeneous {B}ose gas},\ }
 \href{https://link.aps.org/doi/10.1103/PhysRevLett.129.190402} 
 {\bibfield {journal} {\bibinfo  {journal} {Phys. Rev. Lett.}\ }\textbf {\bibinfo {volume}{129}},\ 
 \bibinfo {pages} {190402} (\bibinfo {year} {2022})}\BibitemShut {NoStop}%
\end{thebibliography}


\clearpage

\pagebreak
\twocolumngrid
\begin{center}
\textbf{\large Supplemental Material: \TITLE}
\end{center}
%%%%%%%%%% Prefix a "S" to all equations, figures, tables and reset the counter %%%%%%%%%%
\setcounter{equation}{0}
\setcounter{figure}{0}
\setcounter{table}{0}
\makeatletter
\renewcommand{\theequation}{S\arabic{equation}}
\renewcommand{\thefigure}{S\arabic{figure}}
\renewcommand{\bibnumfmt}[1]{[S#1]}

\Sec{Excitation method}
The excitation potential is created by a pair of coils in an anti-Helmohltz configuration, with one of the coils aligned and the other displaced in $z$ and $y$ with respect to the symmetry axis of the trapping potential. An oscillatory electric current through the coils generates a potential that shifts the trapping center and causes compression to the cloud. Our excitation protocol promotes energy injection predominantly in a momentum scale of the order of the inverse of the Thomas-Fermi radius of the BEC ($R$), $k \sim 1/R \approx 0.25\ \mu\mathrm{m}^{-1}$.
Although our trap is symmetric in $y$ and $z$ directions, our excitation couples energy predominantly in a single direction since we only observe momentum spread along the $z$ direction. As a result, the system coupling is primarily in the dipole mode. The magnitude, frequency, phase, and duration of the electric current passing through the coils can be varied to study different excitation regimes. In all measurements reported in this work, we fixed the oscillation period $\tau$, the excitation time \(t_{\mathrm{exc}} = 6\tau\), and the excitation frequency \(\omega_{\mathrm{exc}} = 2\pi/\tau = 2\pi \times 110\) Hz. We calibrate the potential generated by the excitation coils by turning on the current at a fixed value for a short time $\Delta t$ immediately after releasing the cloud from the trap and then measuring the resulting velocity increase in time of flight $\Delta v = d/t_{\mathrm{ToF}}$, $d$ being the displacement of the center of mass. As a result, we obtain $\Delta U = m 2R d /(\Delta t \ t_{\mathrm{ToF}}$). This potential variation $\Delta U$, expressed in units of the chemical potential in the center of the cloud in equilibrium ($\mu_0$), is what we denote the excitation amplitude $A$ throughout the text, and it has an uncertainty of $\approx 6$\%.

\begin{figure}[htb]
     \centering
     \includegraphics[scale = 1]{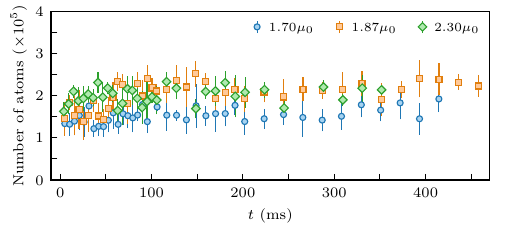}
     \caption{Total number of atoms after excitation for the three amplitudes presented in Fig.~2. It remains constant with a small variation over all considered times. The error bars correspond to the standard deviation of the set of repetitions of the same measurement.}
     \label{fig:Natoms}
\end{figure}

For all excitation amplitudes used, the number of atoms as a function of time is monitored to ensure that we have a system with an approximately constant number of particles. Figure~\ref{fig:Natoms} shows the number of atoms as a function of the time for the three amplitudes shown in Fig.~2. This number remains constant and reveals the type of shot-to-shot fluctuations the experiment presents. For each instance, we repeated the measurement at least six times. The error bars represent the standard deviation with respect to the mean value.

\begin{figure*}[htb]
    \centering
    \includegraphics[scale=1]{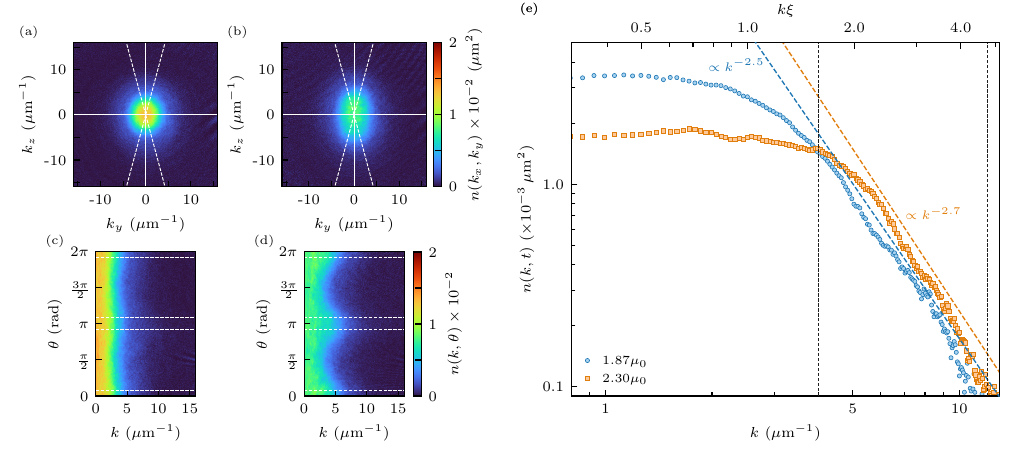}
    \caption{Single shot illustrations of the angular averaging procedure to compute the momentum distributions for the excitation amplitudes (a) $A=1.87\mu_0$ and (b) $2.30\mu_0$, in both cases for $t\approx 100$ ms. The dashed lines denote two $30^{\circ}$ angles centered around the major axis of the expanded cloud enclosing the region where $n(k)$ is calculated. (c),(d) The same information as (a) and (b), but expressed in polar coordinates. (e) Resulting angular averaged momentum distribution curves. The appearance of power-law behavior in the momentum range $4 \ \mu\mathrm{m}^{-1}\leqslant k \leqslant 12\  \mu\mathrm{m}^{-1}$ supports the establishment of a turbulent state.}
    \label{fig:momentum_distribution}
\end{figure*}

\begin{figure*}[htb]
    \centering
    \includegraphics[scale=1]{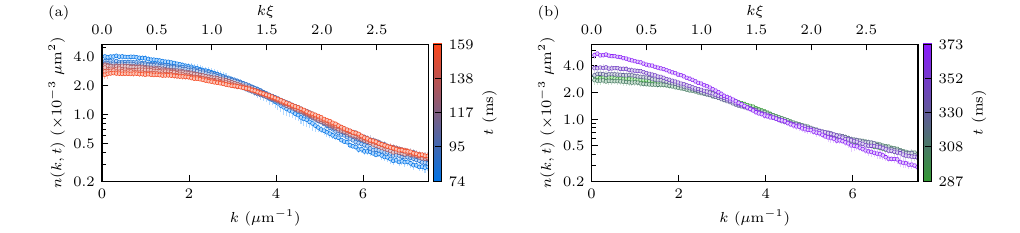}
    \caption{Momentum distributions for an excitation amplitude of $A=1.87 \mu_0$ during (a) the direct and (b) inverse particle cascades. These plots correspond to a larger momentum range than Figs.~4(b) and (e), where it is possible to see the particle transport from low to high momentum during the direct cascade and in the opposite direction for the inverse cascade.}
    \label{fig:larger_k_range}
\end{figure*}

\begin{figure*}[htb]
    \centering
    \includegraphics[scale=1]{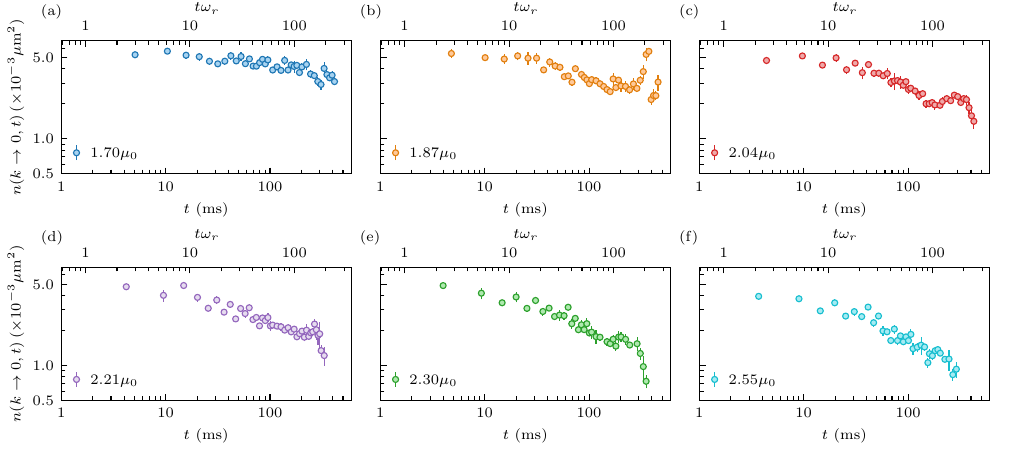}
    \caption{Time evolution of $n(k\rightarrow 0,t)$ for the excitation amplitudes $A=1.70\mu_0$, $1.87\mu_0$, $2.04\mu_0$, $2.21\mu_0$, $2.30\mu_0$, and $2.55 \mu_0$, panels (a)-(f), respectively.}
    \label{fig:nk0_all_amplitudes}
\end{figure*}

\Sec{Measurement of the momentum distributions}
To study the relaxation and thermalization of a system that departed from equilibrium, we measure its time-dependent momentum distribution. We obtain the \textit{in situ} momentum distribution $\tilde{n}(k,t)$ of our BECs via absorption images after the time of flight. The captured image is a two-dimensional (2D) projection of the three-dimensional (3D) atomic density, with the density information integrated over the $x$-axis, allowing for determining the column density in the $yz$ plane. The position coordinates in the atomic density distribution are converted to momentum coordinates $k$ via $r = \hbar t_{\mathrm{ToF}} k/m$, where $k=(k_y^2+k_z^2)^{1/2}$. The momentum distribution is normalized by the total number of atoms, $n(k,t) = \tilde{n}(k,t)/N(t)$, to prevent uncertainties resulting from small fluctuations between different measurements. Atoms are always released from the harmonic trap at a multiple of a quarter of the radial oscillation period, $T_r/4$, i.e., when the \textit{in situ} atomic distribution reflects the momentum distribution, which is then amplified by time-of-flight measurements.

To mitigate the anisotropy effects of the momentum distributions, we only consider momentum regions close to the major axis of the expanded cloud, i.e., our angular averaging is restricted to an aperture of 30$^{\circ}$ around the $z$-axis. Two examples for the excitation amplitudes of $1.87\mu_0$ and $2.30\mu_0$ are illustrated in Fig.~\ref{fig:momentum_distribution}(a) and (b), respectively. Their corresponding angular distributions are shown in Fig.~\ref{fig:momentum_distribution}(c) and (d), where the dashed lines limit the regions where the angular average in $\theta$ is performed. Figure~\ref{fig:momentum_distribution}(e) shows the obtained momentum distributions $n(k,t\approx 100 \text{ ms})$. Both cases display a particle cascade characterized by a power law in the inertial region (indicated by the vertical dashed lines), a signature of the establishment of a turbulent state~\cite{Thompson_2014,navon2016emergence}.

The direct cascade is characterized by particle transport from low- to high-momenta, while the inverse cascade occurs in the opposite direction. Both scenarios are illustrated in Fig.~\ref{fig:larger_k_range}, where we present the time-evolving momentum distributions during both cascades. The values of $n(k \rightarrow 0, t)$ reported throughout the manuscript are the average of all values with $k \leqslant 0.33\ \mu\mathrm{m}^{-1}$. The pixel size in our absorption images corresponds to a momentum value of $0.066\ \mu\mathrm{m}^{-1}$.

The observation of the inverse cascade depends on a delicate balance. If too much energy is injected into the system, the condensate is quickly depleted and thermalized. On the other hand, if not enough energy is injected, the population of the high-momentum modes does not increase significantly, and identifying the inverse cascade is challenging.

Figure~\ref{fig:nk0_all_amplitudes} shows the time evolution of $n(k\rightarrow 0,t)$ for the three excitation amplitudes depicted in Fig.~2 alongside three other excitation amplitudes, $A=2.04\mu_0$, $2.21\mu_0$, and $2.55 \mu_0$. For $A = 1.70\mu_0$, the excitation amplitude is so low that we do not observe characteristics of a turbulent system, such as a power-law behavior in the momentum distribution, nor compelling evidence of repopulation of the low-momentum modes. As the excitation amplitude progressively increases, the system shows evidence of condensate repopulation at $\approx 300$ ms. The most apparent evidence of this phenomenon was measured with an excitation amplitude of $1.87\mu_0$ at $t\approx 370$ ms, which is the one we thoroughly investigate in the manuscript. Using higher excitation amplitudes, where we also observed the repopulation of low momenta, the system fails to reach such high values of $n(k\to 0,t)$. For excitation amplitudes $A \geqslant 2.04\mu_0$, we also observed that when the repopulation occurs, it is always at times close to 300 ms, with a slight advancement relative to lower amplitudes, e.g., for $A=2.04\mu_0$ the peak occurs at $t\approx 300$ ms, while for $A=2.21\mu_0$ at $t\approx 260$ ms.

\Sec{Determining the universal exponents related to the direct cascade}
Although in the main text we focus on the NTFP present in the dynamics of the system subjected to an excitation amplitude of $1.87\mu_0$, we observe the universal scaling described by Eq.~(1) in a total of eight excitation amplitudes in the interval $1.87 \mu_0 \leqslant A \leqslant 2.55 \mu_0$.

\begin{figure}[htb]
    \centering
    \includegraphics[scale=1]{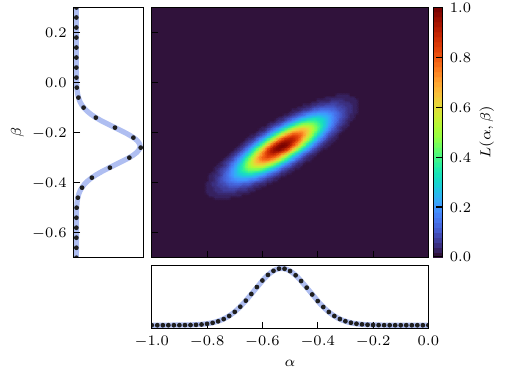}
    \caption{Likelihood function for eight excitation amplitudes in the range $1.87 \mu_0 \leqslant A \leqslant 2.55 \mu_0$. The left and bottom panels correspond to the marginal-likelihood functions. The blue curves are Gaussian fits, from which we obtain the values of the universal exponents and their uncertainty, $\alpha = -0.5(1)$ and $\beta = -0.25(7)$.}
    \label{fig:likelihood}
\end{figure}

We followed the same procedure as in Ref.~\cite{PhysRevA2022} to determine the universal exponents. In summary, we compute a likelihood function which considers all possible values of $\alpha$ and $\beta$ for a given set of momentum distributions in a finite momentum range and at a fixed excitation amplitude, $L_A(\alpha,\beta)$. We combine the results of several excitation amplitudes by computing their product,
\begin{equation}
    L(\alpha,\beta)=\prod_A L_A(\alpha,\beta).
\end{equation}
The values of $\alpha$ and $\beta$, and their uncertainties, are obtained from Gaussian fits to the marginal-likelihood functions,
\begin{eqnarray}
    \label{eq:Lalpha}
    L_\alpha(\alpha)&=&\int d\beta\ L(\alpha,\beta), \\
    \label{eq:Lbeta}
    L_\beta(\beta)&=&\int d\alpha\ L(\alpha,\beta).
\end{eqnarray}
In Fig.~\ref{fig:likelihood}, we present $L(\alpha,\beta)$ and the marginal-likelihood functions in the left and bottom panels.

\begin{figure}[htb]
    \centering
    \includegraphics[scale=1]{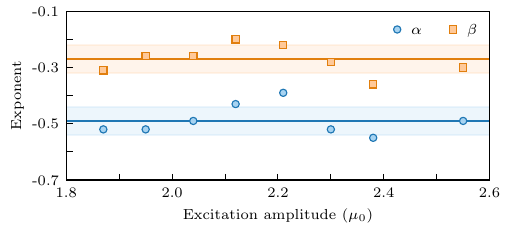}
    \caption{Universal exponents for excitation amplitudes $1.87\mu_0 \leqslant A \leqslant 2.55\mu_0$ obtained with an alternative approach~\cite{madeira2023differential}. The shaded regions indicate one standard deviation.}
    \label{fig:exponents}
\end{figure}

The values of the universal exponents we obtained, $\alpha = -0.5(1)$ and $\beta = -0.25(7)$, are in agreement with the values reported in Ref. \cite{PhysRevA2022}. Additionally, for comparison, we calculated the universal exponents using another approach based on a differential equation introduced in Ref.~\cite{madeira2023differential}. This method analyses the behavior of the momentum distributions in the vicinity of just two momentum values: $k \rightarrow 0$ and $k^{*}$, the latter being the value where the distributions are approximately time-independent. Figure~\ref{fig:exponents} displays the values of $\alpha$ and $\beta$ for amplitudes between $1.87\mu_0$ and $2.55\mu_0$ computed with this alternative approach. By averaging over all amplitudes, we obtained $\alpha = -0.49(5)$ and $\beta = -0.27(5)$, in good agreement with the other adopted approach.

As discussed above, the momentum distributions $n(k,t)$ used throughout this work correspond to 2D projections of the original 3D system. In Ref.~\cite{PhysRevA2022}, it was shown that the projection of a momentum distribution that displays the universal scaling of Eq.~(1) also presents universal dynamical scaling if we assume isotropy. The relation between the exponents obtained through the study of the projected cloud and the ones of the three-dimensional system is
\begin{eqnarray}
    \alpha_{\mathrm{2D}} &=& \alpha_{\mathrm{3D}} - \beta_{\mathrm{3D}},\\
    \beta_{\mathrm{2D}} &=& \beta_{\mathrm{3D}},
\end{eqnarray}
where the sub-indices indicate if the exponent corresponds to the 3D system or its 2D projection. This result indicates that the universal scaling of an isotropic three-dimensional momentum distribution is preserved during the projection process.

\Sec{Determining the universal exponents related to the inverse cascade}
We followed a similar procedure to the one described above, or in Ref.~\cite{PhysRevA2022}, to determine the universal exponents $\lambda$ and $\mu$. Compared to the procedure to extract $\alpha$ and $\beta$, a complication is that we must also determine the value of the blowup time $t_b$. We circumvent this problem by also varying $t_b$, and choosing the value which minimizes the uncertainties in $\lambda$ and $\mu$.

We start by defining the function
\begin{equation}
\chi^2(\lambda,\mu)=\frac{1}{N_t^2}\sum_{t=t_1}^{t_{N_t}}\sum_{t_{\rm ref}=t_1}^{t_{N_t}} \chi^2_{\lambda,\mu}(t,t_{\rm ref}),
\end{equation}
where we the sums run over all time instants $\{t_1,\cdots,t_{N_t}\}$. The function $\chi^2_{\lambda,\mu}(t,t_{\rm ref})$ is given by
\begin{equation}
\label{eq:local}
\chi^2_{\lambda,\mu}(t,t_{\rm ref})=\int_{0}^{k_f} dk \frac{[\tau^\lambda n(\tau^\mu k,t_{\rm ref})-n(k,t)]^2}{\sigma(\tau^\mu k,t_{\rm ref})^2+\sigma(k,t)^2},
\end{equation}
where $\sigma(k,t)$ corresponds to the standard deviation of the mean. The upper limit of the integration $k_f$ was varied, and we found that the exponents are insensitive to variations of $\delta k\approx\unit[1.0]{\mu{m}^{-1}}$ around $k_f=\unit[6.0]{\mu{m}^{-1}}$.

Using the likelihood function
\begin{equation}
L(\lambda,\mu)=\exp\left[-\frac{1}{2}\chi^2(\lambda,\mu)\right],
\end{equation}
the exponents and their uncertainties are determined from a Gaussian fit of the marginal-likelihood functions, the same procedure used in Eqs.~(\ref{eq:Lalpha}) and (\ref{eq:Lbeta}) with $\alpha$ and $\beta$ interchanged by $\lambda$ and $\mu$, respectively.

\begin{figure*}[htb]
    \centering
    \includegraphics[scale=1]{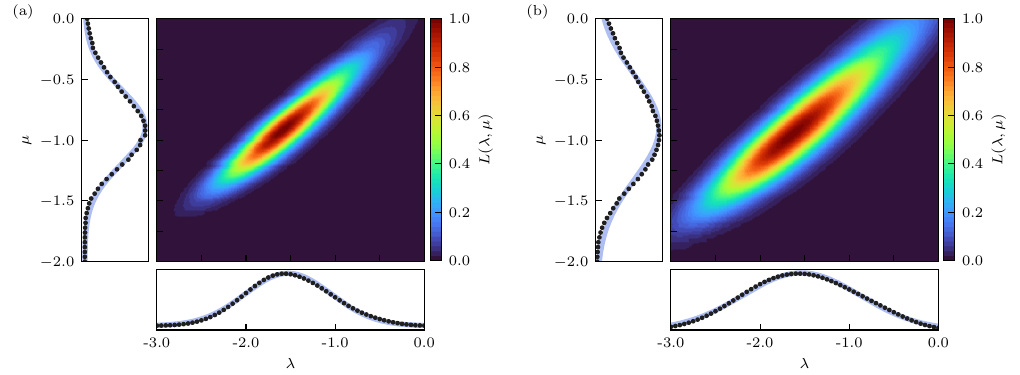}
    \caption{Likelihood functions for an excitation amplitude of $A=1.87 \mu_0$ used to determine the exponents of the dynamical scaling related to the inverse cascade, Eq.~(2). (a) The temporal window $288$ ms $\leqslant t \leqslant 374$ ms yields the exponents reported in the main text, $\lambda=-1.5(5)$ and $\mu=-0.9(3)$. (b) A slightly larger interval, $266$ ms $\leqslant t \leqslant 374$ ms, produces the same exponents with larger uncertainties, $\lambda=-1.5(7)$ and $\mu=-0.9(4)$.}
    \label{fig:likelihood_lambda_mu}
\end{figure*}

In Fig.~\ref{fig:likelihood_lambda_mu}(a), we show the likelihood function and its marginal likelihoods in the bottom and left panels for the case presented in the main text, a universal scaling window of $288$ ms $\leqslant t \leqslant 374$ ms. We obtained $\lambda=-1.5(5)$ and $\mu=-0.9(3)$ for a blowup time of $t_b=520(10)$ ms. If we consider a slightly larger temporal window, $266$ ms $\leqslant t \leqslant 374$ ms, we obtain the likelihoods illustrated in Fig.~\ref{fig:likelihood_lambda_mu}(b).
The exponents are the same as the previous case but with more considerable uncertainties, $\lambda=-1.5(7)$ and $\mu=-0.9(4)$ for $t_b=560(10)$ ms. This indicates that the dynamical scaling region may be more restricted than indicated by only the $k\to 0$ component. The fit shown in Fig.~3 considers the functional form $n(k\to 0,t)\propto (t_{b}-t)^{\lambda}$ where we fixed the value of $\lambda$ at its theoretical estimate, ${\lambda} ={\nu/(2-2\nu)}= -1.46$. We obtained $t_b=530(30)$ ms in accordance with the full scaling procedure considering finite values of the momentum.

\begin{figure*}[htb]
    \centering
    \includegraphics[scale=1]{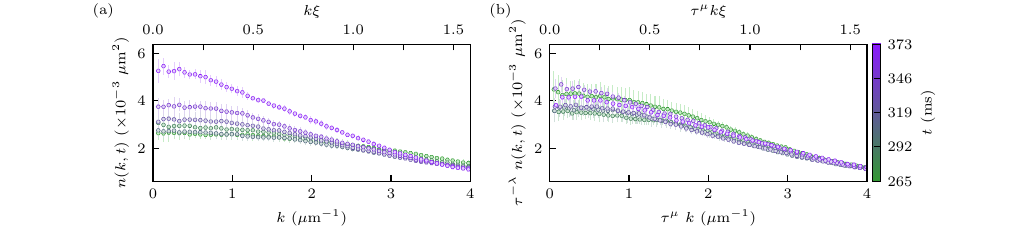}
    \caption{(a) Momentum distributions for an excitation amplitude of $A=1.87 \mu_0$ during $266$ ms $\leqslant t \leqslant 374$ ms, a larger interval than the one considered in the main text, Fig.~4(e). (b) Scaled momentum distributions using Eq.~(2) with $\lambda=-1.5(7)$ and $\mu=-0.9(4)$.}
    \label{fig:scaling_blowup_6}
\end{figure*}

In Fig.~\ref{fig:scaling_blowup_6}, we show the result of the dynamical scaling applied to the larger temporal window, $266$ ms $\leqslant t \leqslant 374$ ms, the equivalent of panel (e) and (f) of Fig.~4. Although it is possible to identify the collapse after the universal scaling procedure, the dispersion is more significant than the case we considered in the main text. This is in agreement with the larger uncertainties in the exponents if we increase the temporal window.

\end{document}